\documentclass[11pt]{article}
\usepackage{graphicx}
\usepackage{latexsym,graphicx}
\usepackage{float}
\usepackage{amsmath}
\usepackage{amsfonts}
\usepackage{amssymb}
\usepackage{epstopdf}
\usepackage{color}
\DeclareGraphicsExtensions{.eps}
\usepackage[left=2.5cm,top=2.5cm,right=2.5cm,bottom=2.5cm]{geometry}

\catcode `\@=11
\catcode `\@=12
\begin{document}
	\begin{center}
		\large{\bf{Exact Cosmology in Myrzakulov Gravity}} \\
		\vspace{5mm}
		\normalsize{ Dinesh Chandra Maurya$^{1}$,  Ratbay Myrzakulov$^{2}$}\\
		\vspace{5mm}
		\normalsize{$^{1}$ Centre for Cosmology, Astrophysics and Space Science, GLA University, Mathura-281 406,
			Uttar Pradesh, India.}\\
		\vspace{5mm}
		\normalsize{$^{2}$ Eurasian International Centre for Theoretical Physics and Department of General \& Theoretical Physics, Eurasian National University, Astana 010008, Kazakhstan.}\\
		\vspace{2mm}
		{$^{1}$Email:dcmaurya563@gmail.com}\\
		\vspace{2mm}
		{$^{2}$Email:rmyrzakulov@gmail.com}\\
	\end{center}
	\vspace{5mm}
	%\date{}
	%\maketitle
	%%%%%%%%%%%%%%%%%%%%%%%%%%%%%%%%%%%%%%%%%%%%%%%%%%%%%%%%%%%%%%%%%%%%%%%%%%%%%%%%%%%%%%%%%%%%
	\begin{abstract}
     In this paper, we have investigated some exact cosmological models in Myrzakulov gravity using a flat Friedmann-Lematre-Robertson-Walker (FLRW) spacetime metric. We have considered the modified Lagrangian function as $F(R,T)=R+\lambda T$, where $R, T$ are respectively the Ricci curvature scalar and the torsion scalar with respect to non-special connection, and $\lambda$ is a model parameter. We have obtained two exact solutions in two different situations for the scale factor $a(t)$. Using this scale factor, we have obtained various geometrical parameters to investigate cosmological properties of the universe. We have obtained the best fit values of model parameters through the MCMC analysis of two types latest observational datasets like $H(z)$ and Pantheon SNe Ia samples, with $1-\sigma, 2-\sigma$ \& $3-\sigma$ regions. We have performed a comparative and relativistic study of geometrical and cosmological parameters. In model-I, we have found that the effective equation of state (EoS) parameter $\omega_{eff}$ varies in the range $-1\le\omega_{eff}\le0$ while in the model-II, it varies as $-1.031\le\omega_{eff}\le0$. We have found that both models are transit phase (decelerating to accelerating) universe with transition redshift in the range $0.6<z_{t}<0.8$ and present age of the universe $t_{0}\approx13.5$ Gyrs.
	\end{abstract}
	\smallskip
	\vspace{5mm}
	%\date{}
	%\maketitle
	\vspace{1cm}
	
	\tableofcontents
	%%%%%%%%%%%%%%%%%%%%%%%%%%%%%%%%%%%%%%%%
	\section{Introduction}
	%%%%%%%%%%%%%%%%%%%%%%%%%%%%%%%%%%%%%%%%
	
	Recent cosmological studies \cite{ref1}-\cite{ref7} suggest that the Universe evolves with two accelerating phases of expansion, one at early times and one at late times. Although the latter can be explained by the presence of a cosmological constant $\Lambda$, the inclusion of some modification was necessary due to the associated theoretical issue, the potential for a dynamical behavior, and particularly the cosmological constant's incapacity to explain the early accelerated phase. One possible approach is to stick with general relativity as the basic theory and change the universe's matter content by adding new fields, like the dark energy sector at later times \cite{ref8,ref9} and/or the inflaton at earlier times \cite{ref10,ref11}. A second option involves altering the gravitational sector itself, creating a theory that, although having general relativity as a specific limit, generally shows more degrees of freedom \cite{ref12,ref13}.\\

    Gravitational modifications are constructed in a variety of methods, each of which modifies a different aspect of general relativity. The braneworld theories \cite{ref14} comes from changing the dimensionality; the $F(R)$ gravity \cite{ref15,ref16}, $F(G)$ gravity \cite{ref17,ref18}, Lovelock theories \cite{ref19,ref20}, etc. come from changing the Einstein-Hilbert Lagrangian; and the Horndeski/Galileon theories \cite{ref21,ref22,ref23} come from adding a scalar field coupled in different ways with curvature. Furthermore, one can create modifications utilizing torsional invariants, as in $F(T)$ gravity \cite{ref24,ref25}, in $F(T, T_{G})$ gravity \cite{ref26}, or in scalar-torsion theories \cite{ref27,ref28}, beginning with the analogous, teleparallel formulation of gravity \cite{ref29,ref30}. Furthermore, a broad class of metric-affine theories \cite{ref31,ref32,ref33} can be built, which includes a generic linear connection structure; alternatively, nonlinear connections can be introduced, as in the case of Finsler and Finsler-like theories \cite{ref34}-\cite{ref39}.\\

    Motivated by these, a theory utilizing a specific but non-special connection could be built from such affinely connected metric theories, and specifically from their Riemann-Cartan subclass \cite{ref40}. This would result in both non-zero curvature and non-zero torsion at the same time, providing the additional degrees of freedom usually required in any gravitational modification \cite{ref41}. As a result, Myrzakulov gravity can produce a good phenomenology that can explain both the early and late universe evolution \cite{ref42}–\cite{ref46}. A recent analysis of the resulting cosmology of such a framework and the computation of the evolution of observable quantities such as the density parameters and the effective dark energy equation-of-state parameter have been published in \cite{ref42}. Through the use of the mini-super-space technique, they have studied the cosmological behavior with a focus on the effect of the connection, expressing the theory as a deformation from both general relativity and its teleparallel counterpart. \cite{ref47} has investigated the observational restrictions on Myrzakulov $F(R,T)$-gravity. In \cite{ref48}–\cite{ref54}, several Metric-Affine Myrzakulov Gravity Theories and their applications are covered.\\

   Recently \cite{ref55} has investigated observational constraints on some Myrzakulov gravity models using a wide range of observational datasets. In \cite{ref56}, we have investigated some exact solutions in Myrzakulov $F(R,T)$ gravity with observational constraints. Recently we have investigated some exact cosmological models in different scenarios in various modified gravity theories \cite{ref57}-\cite{ref63}. Motivated by above discussions, in this paper, we find some new exact cosmological models in Myrzakulov gravity.\\

   This paper is organized as follows. Section 1 is introductory. A brief introduction of Myrzakulov gravity is given in Section 2. Field equations and their some exact solutions are given in Section 3. We perform some observational constraints on different parameters using latest datasets in Section 4. Result discussions are given in Section 5. Finally, conclusions are given in last Section 6.

	%%%%%%%%%%%%%%%%%%%%%%%%%%%%%%%%%%%%%%%%%%%%%%%%%%%%%%%%%%%%%%%
	\section{Myrzakulov Gravity}
	%%%%%%%%%%%%%%%%%%%%%%%%%%%%%%%%%%%%%%%%%%%%%%%%%%%%%%%%%%%%%%%
	
	In the present section, we organize a brief review of Myrzakulov gravity, also called as $F(R,T)$ gravity with $R$ as curvature and $T$ known as torsion \cite{ref41,ref42}. The key factor of this gravity is the choice of non-special connection. Specifically, it is established that enforcing a generic connection ${\omega^{\alpha}}_{\beta\gamma}$ one defines the torsion tensor and the curvature as \cite{ref26}
	\begin{equation}\label{eq1}
		{R^{\alpha}}_{\beta ij}={\omega^{\alpha}}_{\beta j,i}-{\omega^{\alpha}}_{\beta i,j}+{\omega^{\alpha}}_{\gamma i}{\omega^{\gamma}}_{\beta j}-{\omega^{\alpha}}_{\gamma j}{\omega^{\gamma}}_{\beta i}
	\end{equation}
	\begin{equation}\label{eq2}
		{T^{\alpha}}_{ij}={e^{\alpha}}_{j,i}-{e^{\alpha}}_{i,j}+{\omega^{\alpha}}_{\beta i}{e^{\beta}}_{j}-{\omega^{\alpha}}_{\beta j}{e^{\beta}}_{i}
	\end{equation}
	where ${e_{\alpha}}^{i}\partial_{i}$ is the tetrad field associated with the metric through $g_{ij}=\eta_{\alpha\beta}{e^{\alpha}}_{i}{e^{\beta}}_{j}$, where $\eta_{\alpha\beta}=diag(1, -1, -1, -1)$, where the Latin, Greek indices signify the coordinate and tangent space, respectively, and where the comma indicates differentiation.\\	
	There are numerous possibilities for connections. The only connection that provides vanishing torsion is the Levi-Civita $\Gamma_{\alpha\beta\gamma}$. Henceforth, we will refer to the curvature (Riemann) tensor as ${R^{(LC)\alpha}}_{\beta ij}={\Gamma^{\alpha}}_{\beta j,i}-{\Gamma^{\alpha}}_{\beta i,j}+{\Gamma^{\alpha}}_{\gamma i}{\Gamma^{\gamma}}_{\beta j}-{\Gamma^{\alpha}}_{\gamma j}{\Gamma^{\gamma}}_{\beta i}$. The Weitzenb\"{o}ck connection, on the other hand, can be used to obtain the following: ${T^{(W)k}}_{ij}={W^{k}}_{ji}-{W^{k}}_{ij}$. This connection is curvatureless and yields only torsion. Quantities corresponding to ${W^{k}}_{ij}$ are denoted by the label ``W". It is inferred from the preceding that the Levi-Civita connection's associated Ricci scalar is
	\begin{equation}\label{eq3}
		R^{(LC)}=\eta^{\alpha\beta}{e_{\alpha}}^{i}{e_{\beta}}^{j}\left[{\Gamma^{k}}_{ij,k}-{\Gamma^{k}}_{ik,j}+{\Gamma^{l}}_{ij}{\Gamma^{k}}_{kl}-{\Gamma^{l}}_{ik}{\Gamma^{k}}_{jl}\right]
	\end{equation}
	while the torsion scalar corresponding to the Weitzenb\"{o}ck connection is
	\begin{equation}\label{eq4}
		T^{(W)}=\frac{1}{4}\left(W^{ikj}-W^{ijk}\right)\left(W_{ikj}-W_{ijk}\right)+\frac{1}{2}\left(W^{ikj}-W^{ijk}\right)\left(W_{kij}-W_{kji}\right)-\left({W_{j}}^{ij}-{W_{j}}^{ji}\right)\left({W^{k}}_{ik}-{W^{k}}_{ki}\right)
	\end{equation}
	The first is utilized in the Lagrangian of General Relativity and in all curvature-based modified gravities, such as in $F(R)$ gravity \cite{ref15}, whereas the second is utilized in the Lagrangian of the teleparallel equivalent of General Relativity and in all torsion-modified gravities, such as in $F(T)$ gravity \cite{ref13}.\\	
	A non-special connection with simultaneous non-zero curvature and torsion is used in Myrzakulov gravity \cite{ref42}. Since general relativity and its teleparallel counterpart, which both have two degrees of freedom corresponding to the massless graviton, do not have extra degrees of freedom, the resulting theory will generally have them, even if the imposed Lagrangian is simple. Here, we consider the action as
	\begin{equation}\label{eq5}
		I=\int{d^{4}x~e\left[\frac{1}{2\kappa^{2}}F(R,T)+L_{m}\right]}
	\end{equation}
	where the gravitational constant is $\kappa^{2}=8\pi G$, the matter Lagrangian $L_{m}$ has also been introduced for completeness, and $e=\det(e^{\alpha}_{i})=\sqrt{-g}$. Take note that the arbitrary function $F(R, T)$ has two variables, $R$ and $T$, which represent the curvature and torsion scalars, respectively, corresponding to the non-special connection employed \cite{ref26}.	
	\begin{equation}\label{eq6}
		T=\frac{1}{4}T^{ijk}T_{ijk}+\frac{1}{2}T^{ijk}T_{kji}-{T_{j}}^{ji}{T^{k}}_{ki},
	\end{equation}
	\begin{equation}\label{eq7}
		R=R^{(LC)}+T-2{{T_{j}}^{ji}}_{;i}
	\end{equation}
	where the covariant differentiation with regard to the Levi-Civita connection is indicated by the symbol (;). So, while $R$ depends on the tetrad and its first derivative as well as the connection and its first derivative, it also depends on the second tetrad derivative because of the final term in \eqref{eq7}. In contrast, $T$ depends on the tetrad field, its first derivative, and the connection. Therefore, we can finally write using Eqs.~\eqref{eq4}, \eqref{eq6}, and \eqref{eq7}.
	\begin{equation}\label{eq8}
		R=R^{(LC)}+u,
	\end{equation}
	\begin{equation}\label{eq9}
		T=T^{(W)}+v,
	\end{equation}
	The scalar $u$ is contingent upon the tetrad, its first and second derivatives, the connection, and its first derivative, whereas the scalar $v$ is contingent upon the tetrad, its first derivative, and the connection.\\	
	The effect of the specific but non-special imposed connection is quantified by the numbers $u$ and $v$. If this link turns into the Levi-Civita connection, then $u = 0$ and $v = -T^{(W)}$. This makes the above theory the standard $F(R)$ gravity, which under $F(R) = R$ coincides with general relativity. Conversely, if the connection is the Weitzenb\"{o}ck one, then $v = 0$ and $u = -R^{(LC)}$, and so the theory corresponds with $F(T)$ gravity. For $F(T) = T$, this gravity thus becomes the teleparallel counterpart of general relativity.
	
	%%%%%%%%%%%%%%%%%%%%%%%%%%%%%%%%%%%%%%%%%%%%%%%%%%%%%%%%%%%%%%%%%%%%%%%%%%
	\section{Field Equations and Solutions}
	%%%%%%%%%%%%%%%%%%%%%%%%%%%%%%%%%%%%%%%%%%%%%%%%%%%%%%%%%%%%%%%%%%%%%%%%%
	
	We put all of this to a cosmological framework and get the corresponding field equations that govern the evolution of the universe \cite{ref42}. We consider the Friedmann-Robertson-Walker (FRW) flat Friedmann homogeneous and isotropic geometry spacetime metric as	
	\begin{equation}\label{eq10}
		ds^{2}=dt^{2}-a^{2}(t)\delta_{\mu\nu}dx^{\mu}dx^{\nu},
	\end{equation}
	which is equivalent to the tetrad $e^{\alpha}_{i}=diag[1, a(t), a(t), a(t)]$, in which the scale factor is denoted by $a(t)$. The torsion and curvature scalar $T^{(W)}=-6\left(\frac{\dot{a}}{a}\right)^{2}$ and $R^{(LC)}=6\left(\frac{\ddot{a}}{a}+\left(\frac{\dot{a}}{a}\right)^{2}\right)$ are easily found in this situation, respectively. We further consider the conventional substitution $L_{m}=-\rho_{m}(a)$ \cite{ref64,ref65,ref66}. Ultimately, in light of the previous conversation regarding the interdependence of $u$ and $v$, we consistently enforce that $u=u(a, \dot{a}, \ddot{a})$ and $v=v(a, \dot{a})$.\\	
	We aim to investigate the cosmic behavior resulting just from the non-special connection in Myrzakulov gravity in this work. Since the coupling coefficient of $R$ can be absorbed into $\kappa^{2}$, we remove it. Instead, we concentrate on the simplest case, where the involved arbitrary function is trivial, which is $F(R, T) = R + \lambda T$. Be aware that the Lagrangian does not explicitly take into account a cosmological constant term. By incorporating the aforementioned spacetime metric \eqref{eq10} into the action \eqref{eq5}, we obtain $I=\int{Ldt}$ for this Lagrangian selection \cite{ref55} as
	\begin{equation}\label{eq11}
		L=\frac{3}{\kappa^{2}}[\lambda+1]a\dot{a}^{2}-\frac{a^{3}}{2\kappa^{2}}[u(a, \dot{a}, \ddot{a})+\lambda v(a, \dot{a})]+a^{3}\rho_{m}(a).
	\end{equation}
	Taking the Hamiltonian constraint $H=\dot{a}\left[\frac{\partial L}{\partial \dot{a}}-\frac{\partial}{\partial t}\frac{\partial L}{\partial\ddot{a}}\right]+\ddot{a}\frac{\partial L}{\partial\ddot{a}}-L=0$ and extracting the equations of motion for $a(t)$, we have the Friedmann equations as \cite{ref55}
	\begin{equation}\label{eq12}
		3(1+\lambda)H^{2}-\frac{1}{2}[Ha(u_{\dot{a}}+\lambda v_{\dot{a}})-(u+\lambda v)+au_{\ddot{a}}(\dot{H}-2H^{2})]=\kappa^{2}\rho_{m}
	\end{equation}
	\begin{multline}\label{eq13}
		(1+\lambda)(2\dot{H}+3H^{2})-\frac{1}{6}\left[3Ha(u_{\dot{a}}+\lambda v_{\dot{a}})-3(u+\lambda v)\right. \\\left.-a(u_{a}+\lambda v_{a}-\dot{u}_{\dot{a}}-\lambda\dot{v}_{\dot{a}}) -3a(\dot{H}+3H^{2})u_{\ddot{a}}-6Ha\dot{u}_{\ddot{a}}-a\ddot{u}_{\ddot{a}}\right]=-\kappa^{2}p_{m}
	\end{multline}
	\begin{equation}\label{eq14}
		\dot{\rho}_{m}+3H(\rho_{m}+p_{m})=0
	\end{equation}
    where $H$ is the Hubble parameter defined by $H=\frac{\dot{a}}{a}$, $p_{m}$ denotes the matter pressure, and the subscripts $a, \dot{a}, \ddot{a}$ indicate the partial derivatives with respect to this arguments.\\
	Now, we find the solution of above field equations for two different choices of $u$ and $v$, and we investigate these in next two sub-sections. As per definitions of $u$ and $v$, in first model, we choice $u=k_{1}H-k_{2}$, $v=k_{3}aH-k_{4}$, and in second model, we consider $u=k_{1}(\dot{H}+H^{2})-k_{2}$, $v=k_{3}aH-k_{4}$ with $k_{i}$'s $i=1, 2, 3, 4$ constants,.
	%%%%%%%%%%%%%%%%%%%%%%%%%%%%%%%%%%%%%%%%%%%%%%%%%%%%%%%%%%%%%%%
	\subsection{Model-I}
	%%%%%%%%%%%%%%%%%%%%%%%%%%%%%%%%%%%%%%%%%%%%%%%%%%%%%%%%%%%%%%%
	
	In this model, we choose $u=k_{1}H-k_{2}$ and $v=k_{3}aH-k_{4}$ with $k_{i}$'s $i=1, 2, 3, 4$ constants and $H=\frac{\dot{a}}{a}$, then the above field equations \eqref{eq12} \& \eqref{eq13} become
		\begin{equation}\label{eq15}
		3(1+\lambda)H^{2}-\frac{1}{2}(k_{2}+\lambda k_{4})=\kappa^{2}\rho_{m}
	\end{equation}
	\begin{equation}\label{eq16}
		(1+\lambda)(2\dot{H}+3H^{2})-\frac{1}{2}(k_{2}+\lambda k_{4})=-\kappa^{2}p_{m}
	\end{equation}
	Taking the non-relativistic matter pressure $p_{m}\approx0$ in \eqref{eq16} and rewrite the Eq.~\eqref{eq16} as
	\begin{equation}\label{eq17}
		2\frac{\ddot{a}}{a}+\left(\frac{\dot{a}}{a}\right)^{2}-\frac{k_{2}+\lambda k_{4}}{2(1+\lambda)}=0,~~~~\lambda\ne-1.
	\end{equation}
	Solving the Eq.~\eqref{eq17} for the scale factor $a(t)$, we get
	\begin{equation}\label{eq18}
		a(t)=\left[\frac{3\sqrt{3}c_{2}e^{\frac{\sqrt{3n_{1}}}{2}t}-\sqrt{3}c_{1}e^{-\frac{\sqrt{3n_{1}}}{2}t}}{6\sqrt{n_{1}}}\right]^{\frac{2}{3}},~~~~n_{1}>0,
	\end{equation}
	where $c_{1}, c_{2}$ are arbitrary constants and $n_{1}=\frac{k_{2}+\lambda k_{4}}{2(1+\lambda)}$, $\lambda\ne-1$. Without loss of generality, we choose $c_{1}=k\sqrt{3}, c_{2}=\frac{k}{\sqrt{3}}$ so that we can put the scale factor in the following simplified form (hyperbolic expansion law cosmology)
	\begin{equation}\label{eq19}
		a(t)=\left[\frac{k}{\sqrt{n_{1}}}\sinh{\left(\frac{\sqrt{3n_{1}}}{2}t\right)}\right]^{\frac{2}{3}},~~~~n_{1}>0
	\end{equation}
	Alternatively, if we choose $c_{1}=0$, then we find the exponential expansion law cosmology as $a(t)=\frac{\sqrt{3}c_{2}}{2\sqrt{n_{1}}}e^{\sqrt{\frac{n_{1}}{3}}t}$ which gives a constant deceleration parameter $q$ that reveals either decelerating or accelerating expanding universe, but we seeking a transit phase (decelerating-accelerating) expanding universe model. Therefore, we consider the first choice of $c_{1}, c_{2}$ to investigate the model. Thus using the scale factor mentioned in \eqref{eq19}, we derived the Hubble parameter $H=\frac{\dot{a}}{a}$ and deceleration parameter $q=-\frac{a\ddot{a}}{\dot{a}^{2}}$, respectively as
	\begin{equation}\label{eq20}
		H(t)=\sqrt{\dfrac{n_{1}}{3}}\coth{\left(\frac{\sqrt{3n_{1}}}{2}t\right)}
	\end{equation}
	\begin{equation}\label{eq21}
		q(t)=-1+\frac{3}{2}sech^{2}{\left(\frac{\sqrt{3n_{1}}}{2}t\right)}
	\end{equation}
	Now we use the relationship $a(t)=a_{0}(1+z)^{-1}$ \cite{ref8}, in \eqref{eq20}, \eqref{eq21}, with present value of scale factor $a_{0}=1$ in standard convention, and $z$ as the redshift whose positive values show the early evolution of the universe while the negative values of $z$ reveals the future predictions, and $z=0$ represent the present stage of the universe, we obtain the Hubble parameter $H(z)$ and deceleration parameter $q(z)$ as
	\begin{equation}\label{eq22}
		H(z)=\frac{1}{\sqrt{3}}\sqrt{k^{2}(1+z)^{3}+\frac{k_{2}+\lambda k_{4}}{2(1+\lambda)}}
	\end{equation}
	\begin{equation}\label{eq23}
		q(z)=-1+\frac{3}{2}\frac{k^{2}(1+z)^{3}}{k^{2}(1+z)^{3}+\frac{k_{2}+\lambda k_{4}}{2(1+\lambda)}}
	\end{equation}
	Now we define two more geometrical parameters, proposed in \cite{ref67}, called as statefinder parameters $r, s$, which reveals the geometrical evolution of universe and different stages of dark energy models \cite{ref67,ref68,ref69}. These parameters are defined in terms of scale factor as
	\begin{equation}\label{eq24}
		r=\frac{\dddot{a}}{aH^{3}},~~~~~~s=\frac{r-1}{3(q-\frac{1}{2})}
	\end{equation}
	Using the scale factor \eqref{eq19} in \eqref{eq24}, we obtain the statefinder diagnostic parameters $r(t)$ and $s(t)$ as below
	\begin{equation}\label{eq25}
		r(t)=1-3~sech^{2}{\left(\frac{\sqrt{3n_{1}}}{2}t\right)}
	\end{equation}
	\begin{equation}\label{eq26}
		s(t)=\frac{2~sech^{2}{\left(\frac{\sqrt{3n_{1}}}{2}t\right)}}{3\left[1-sech^{2}{\left(\frac{\sqrt{3n_{1}}}{2}t\right)}\right]}
	\end{equation}
	Now we define the effective equation of state (EoS) parameter $\omega_{eff}$ by comparing the Eqs.~\eqref{eq15} \& \eqref{eq16} to the standard Friedmann equations in a flat spacetime universe, as
	\begin{equation}\label{eq27}
		\omega_{eff}=\frac{p_{eff}}{\rho_{eff}}
	\end{equation}
	or
	\begin{equation}\label{eq28}
		\omega_{eff}=-1+\frac{2n_{1}(\kappa^{2}\rho_{m0}-\lambda k^{2})}{2n_{1}(\kappa^{2}\rho_{m0}-\lambda k^{2})+(k_{2}+\lambda k_{4}-2n_{1}\lambda)k^{2}\sinh^{2}{\left(\frac{\sqrt{3n_{1}}}{2}t\right)}}
	\end{equation}
	Also, from Eq.~\eqref{eq15}, we can derive total energy density parameter as
	\begin{equation}\label{eq29}
		\Omega_{m}+\Omega_{F}=1,
	\end{equation}
	where
	\begin{equation}\label{eq30}
		\Omega_{m}=\frac{\kappa^{2}\rho_{m}}{3H^{2}},~~~~\Omega_{F}=\frac{k_{2}+\lambda k_{4}}{6H^{2}}-\lambda.
	\end{equation}
	called respectively as matter energy density parameter $\Omega_{m}$ and dark energy density parameter $\Omega_{F}$ due to $F(R,T)$ gravity function.\\
	
	In the field equations \eqref{eq15} \& \eqref{eq16}, for $\lambda=0$, one can obtain the Einstein's field equations in GR with cosmological constant as $\Lambda=\frac{k_{2}}{2}$. And in this case, $\Omega_{F}=\frac{k_{2}}{6H^{2}}=\frac{\Lambda}{3H^{2}}=\Omega_{\Lambda}$. Also, for $\lambda\ne0$, we can obtain varying $\Lambda$-term as a function of Hubble parameter $H$, as $\Lambda(H)=\frac{1}{2}(k_{2}-6\lambda H^{2})$.
	
	%%%%%%%%%%%%%%%%%%%%%%%%%%%%%%%%%%%%%%%%%%%%%%%%%%%%%%%%%%%%%%%
	\subsection{Model-II}
	%%%%%%%%%%%%%%%%%%%%%%%%%%%%%%%%%%%%%%%%%%%%%%%%%%%%%%%%%%%%%%%
	
	In this model, we choose $u=k_{1}(\dot{H}+H^{2})-k_{2}$ and $v=k_{3}aH-k_{4}$ with $k_{i}$'s $i=1, 2, 3, 4$ constants, and $\dot{H}+H^{2}=\frac{\ddot{a}}{a}$, in Eqs.~\eqref{eq12} \& \eqref{eq13}, we obtain the following simplified field equations
		\begin{equation}\label{eq31}
		\frac{3}{2}(2+2\lambda+k_{1})H^{2}-\frac{1}{2}(k_{2}+\lambda k_{4})=\kappa^{2}\rho_{m}
	\end{equation}
	\begin{equation}\label{eq32}
		\frac{1}{3}(3+k_{1}+3\lambda)(2\dot{H}+3H^{2})-\frac{1}{2}(k_{2}+\lambda k_{4})=-\kappa^{2}p_{m}
	\end{equation}
		Taking the non-relativistic matter pressure $p_{m}\approx0$ in Eq.~\eqref{eq32}, we rewrite it as 
	\begin{equation}\label{eq33}
		2\frac{\ddot{a}}{a}+\left(\frac{\dot{a}}{a}\right)^{2}-\frac{3(k_{2}+\lambda k_{4})}{2(3+k_{1}+3\lambda)}=0
	\end{equation}
	Solving Eq.~\eqref{eq33} for scale factor $a(t)$, we get
	\begin{equation}\label{eq34}
		a(t)=\left[\frac{3\sqrt{3}c_{4}e^{\frac{\sqrt{3n_{2}}}{2}t}-\sqrt{3}c_{3}e^{-\frac{\sqrt{3n_{2}}}{2}t}}{6\sqrt{n_{2}}}\right]^{\frac{2}{3}},~~~~n_{2}>0
	\end{equation}
	where $c_{3}, c_{4}$ are arbitrary constants and $n_{2}=\frac{3(k_{2}+\lambda k_{4})}{2(3+k_{1}+3\lambda)}$. Without loss of generality, we choose $c_{3}=k\sqrt{3}, c_{4}=\frac{k}{\sqrt{3}}$ so that we can write the scale factor as the hyperbolic expansion law
	\begin{equation}\label{eq35}
		a(t)=\left[\frac{k}{\sqrt{n_{2}}}\sinh{\left(\frac{\sqrt{3n_{2}}}{2}t\right)}\right]^{\frac{2}{3}}
	\end{equation}
	Using this scale factor as in Eq.~\eqref{eq35}, we derive the Hubble parameter $H(t)$ and deceleration parameter $q(t)$ as given below:
	\begin{equation}\label{eq36}
		H(t)=\sqrt{\dfrac{n_{2}}{3}}\coth{\left(\frac{\sqrt{3n_{2}}}{2}t\right)}
	\end{equation}
	\begin{equation}\label{eq37}
		q(t)=-1+\frac{3}{2}sech^{2}{\left(\frac{\sqrt{3n_{2}}}{2}t\right)}
	\end{equation}
	Again, we can express these in terms of redshift $z$ as
		\begin{equation}\label{eq38}
		H(z)=\frac{1}{\sqrt{3}}\sqrt{k^{2}(1+z)^{3}+\frac{3(k_{2}+\lambda k_{4})}{2(3+k_{1}+3\lambda)}}
	\end{equation}
	\begin{equation}\label{eq39}
		q(z)=-1+\frac{3}{2}\frac{k^{2}(1+z)^{3}}{k^{2}(1+z)^{3}+\frac{3(k_{2}+\lambda k_{4})}{2(3+k_{1}+3\lambda)}}
	\end{equation}
	For model-II, the statefinder diagnostic parameters $r(t)$ and $s(t)$ are derived as
		\begin{equation}\label{eq40}
		r(t)=1-3~sech^{2}{\left(\frac{\sqrt{3n_{2}}}{2}t\right)}
	\end{equation}
	\begin{equation}\label{eq41}
		s(t)=\frac{2~sech^{2}{\left(\frac{\sqrt{3n_{2}}}{2}t\right)}}{3\left[1-sech^{2}{\left(\frac{\sqrt{3n_{2}}}{2}t\right)}\right]}
	\end{equation}
	The effective EoS parameter for model-II is derived as
	\begin{equation}\label{eq42}
		\omega_{eff}=-1+\frac{3n_{2}(2\kappa^{2}\rho_{m0}-k_{1}k^{2}-2\lambda k^{2})-n_{2}k_{1}k^{2}\sinh^{2}{\left(\frac{\sqrt{3n_{2}}}{2}t\right)}}{3n_{2}(2\kappa^{2}\rho_{m0}-k_{1}k^{2}-2\lambda k^{2})-3k^{2}(2n_{2}\lambda+n_{2}k_{1}-k_{2}-k_{4}\lambda)\sinh^{2}{\left(\frac{\sqrt{3n_{2}}}{2}t\right)}}
	\end{equation}
		Also, from Eq.~\eqref{eq31}, we can derive total energy density parameter for model-II, as
	\begin{equation}\label{eq43}
		\Omega_{m}+\Omega_{F}=1,
	\end{equation}
	where
	\begin{equation}\label{eq44}
		\Omega_{m}=\frac{\kappa^{2}\rho_{m}}{3H^{2}},~~~~\Omega_{F}=\frac{k_{2}+\lambda k_{4}}{6H^{2}}-\lambda-\frac{1}{2}k_{1}.
	\end{equation}
	Now, we can find the Original Einstein's field equations with cosmological constant $\Lambda$-term in GR, by substituting $\lambda=0, k_{1}=0$, with $\Lambda=\frac{k_{2}}{2}$ otherwise $\Lambda(H)=\frac{1}{2}[k_{2}-(3k_{1}+6\lambda) H^{2}]$.
	%%%%%%%%%%%%%%%%%%%%%%%%%%%%%%%%%%%%%%%%%%%%%%%%%%%%%%%%%%%%%%%%%
	\section{Observational Constraints}
	%%%%%%%%%%%%%%%%%%%%%%%%%%%%%%%%%%%%%%%%%%%%%%%%%%%%%%%%%%%%%%%%%
	In this section, we make observational constraints on the model parameters with observational datasets in our derived model. For this, we use the freely available emcee program at \cite{ref70}, to perform an MCMC (Monte Carlo Markov Chain) analysis so that we can compare our derived model with observational datasets. By varying the parameter values in a probable range of priors and analysis of the parameter space posteriors, the MCMC sampler constraints the cosmological and model parameters.	

	%%%%%%%%%%%%%%%%%%%%%%%%%%%%%%%%%%%%%%%%%%%%%%%%%%%%%%%%%%%%
	\subsection{Hubble function $H(z)$}
	%%%%%%%%%%%%%%%%%%%%%%%%%%%%%%%%%%%%%%%%%%%%%%%%%%%%%%%%%%%%
	The Hubble parameter is one of the most important cosmological parameter for the study of evolution of the universe for both theoretical and observational cosmologists. Due to the availability of observed values of Hubble datasets $H(z)$ with redshift $z$, we first compare our derived Hubble function from the field equations with observed values of $H(z)$ through MCMC analysis, to find the best fit values of model parameters with error bars. For this, we consider $32$ observed Hubble $H(z)$ datasets from \cite{ref71}-\cite{ref72} with errors. We use the following $\chi^{2}$-test formula in our analysis
		\begin{equation}\nonumber
		\chi^{2}(\phi)=\sum_{i=1}^{i=N}\frac{[(H_{ob})_{i}-(H_{th})_{i}]^{2}}{\sigma_{i}^{2}}
	\end{equation}
	Where $N$ denotes the total amount of data, $H_{ob},~H_{th}$, respectively, the observed and hypothesized datasets of $H(z)$ and standard deviations are displayed by $\sigma_{i}$. Here for the Model-I $\phi=(k, k_{2}, k_{4}, \lambda)$ and for the Model-II $\phi=(k, k_{1}, k_{2}, k_{4}, \lambda)$.
	%%%%%%%%%%%%%%%%%%%%%%%%%%%%%%%%%%%%%%%%%%%%%%%%%%%%%%%%%%%%
	%%%%%%%%%%%%%%%%%%%%%%%%%%%%%%%%%%%%% Figure 1
	%%%%%%%%%%%%%%%%%%%%%%%%%%%%%%%%%%%%%%%%%%%%%%%%%%%%%%%%%%%%
	\begin{figure}[H]
		\centering
		\includegraphics[width=10cm,height=10cm,angle=0]{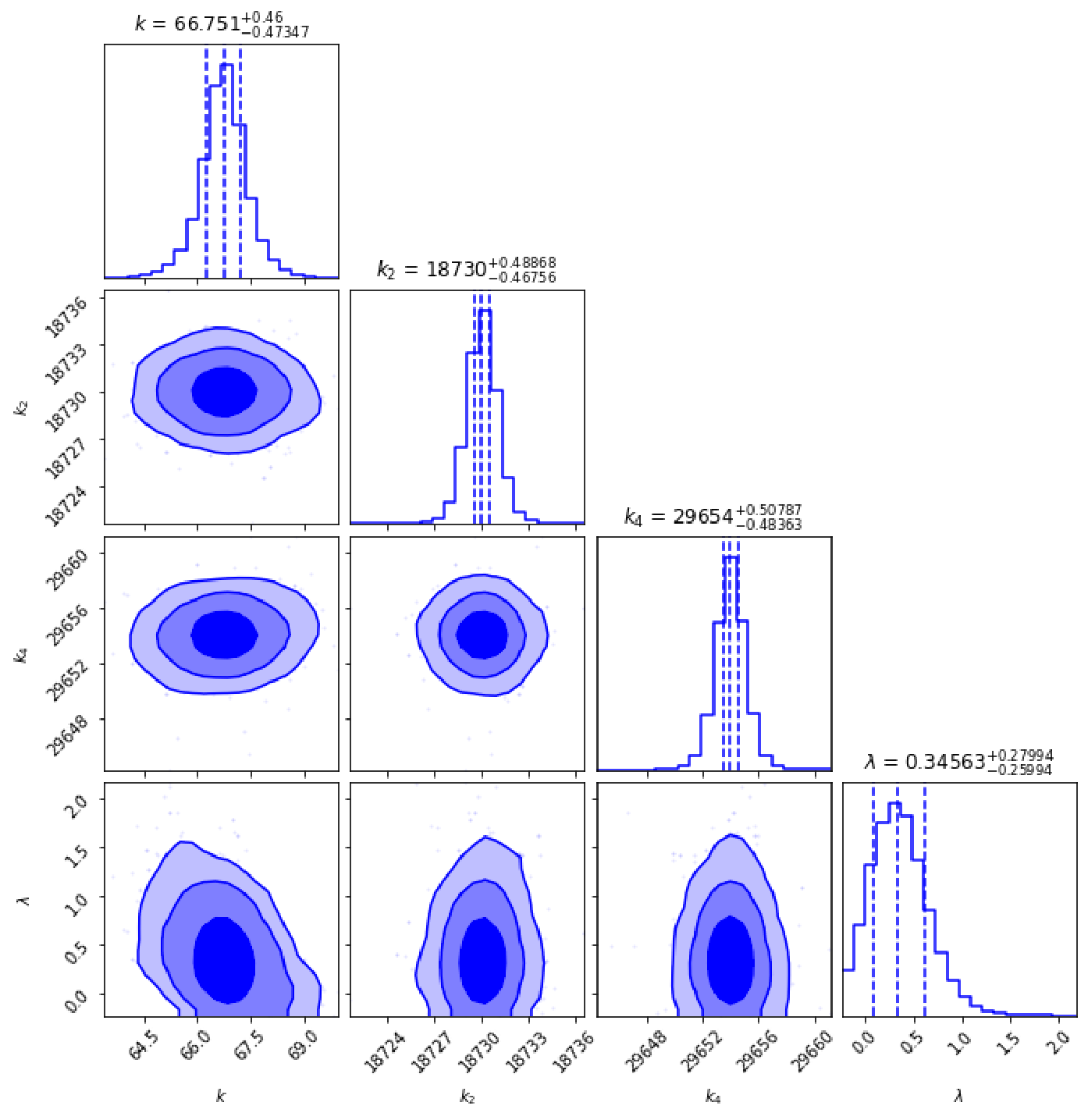}
		\caption{The contour plots of $k, k_{2}, k_{4}, \lambda$ at $1-\sigma, 2-\sigma$ and $3-\sigma$ confidence level in MCMC analysis of $H(z)$ datasets.}
	\end{figure}
	%%%%%%%%%%%%%%%%%%%%%%%%%%%%%%%%%%%%%%%%%%%%%%%%%%%%%%%%%%%%%%%%%%
			%%%%%%%%%%%%%%%%%%%%%%%%%%%%%%%%%%%%%%%%%%%%%%%%%%%%%%%%%%%%%%%%%%%%%	
	\begin{table}[H]
		\centering
		\begin{tabular}{|c|c|c|c|}
			\hline
			% after \\: \hline or \cline{col1-col2} \cline{col3-col4} ...
			
			Parameter       & Prior                    & Model-I  & Model-II   \\
			\hline
			$k$             & $(50, 100)$              & $66.751_{-0.47347}^{+0.46}$      & $66.746_{-0.24954}^{+0.25992}$ \\
			$k_{1}$         & $(-1.0, 2.5)$            & $-$                              & $0.19416_{-0.22047}^{+0.21862}$ \\
			$k_{2}$         & $(10^{4}, 2\times10^{4})$& $18730_{-0.46756}^{+0.48868}$      & $18930_{-0.24383}^{+0.23751}$ \\
			$k_{4}$         & $(10^{4}, 3\times10^{4})$& $29654_{-0.48363}^{+0.50787}$      & $29843_{-0.2494}^{+0.25406}$ \\
			$\lambda$       & $(-0.7, 3.0)$            & $0.34563_{-0.25994}^{+0.27994}$    & $0.20855_{-0.15423}^{+0.21128}$ \\
			\hline
		\end{tabular}
		\caption{The MCMC Results in $H(z)$ datasets analysis.}\label{T1}
	\end{table}
	%%%%%%%%%%%%%%%%%%%%%%%%%%%%%%%%%%%%%%%%%%%%%%%%%%%%%%%%%%%%%%%%%%%%%%%%%%%%
	%%%%%%%%%%%%%%%%%%%%%%%%%%%%%%%%%%%%%%%%%%%%%%%%%%%%%%%%%%%%
	%%%%%%%%%%%%%%%%%%%%%%%%%%%%%%%%%%%%% Figure 2
	%%%%%%%%%%%%%%%%%%%%%%%%%%%%%%%%%%%%%%%%%%%%%%%%%%%%%%%%%%%%
	\begin{figure}[H]
		\centering
		\includegraphics[width=10cm,height=10cm,angle=0]{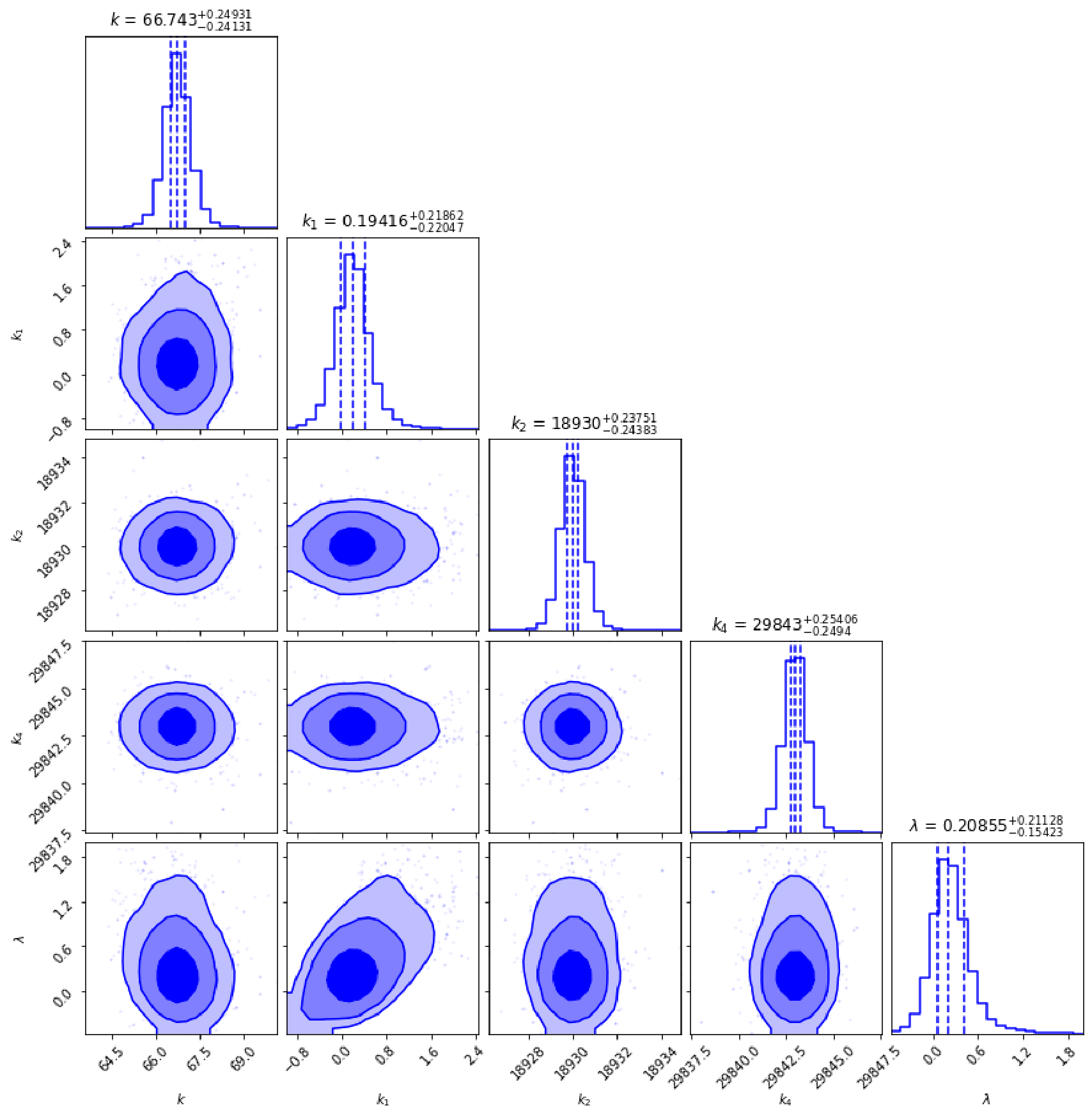}
		\caption{The contour plots of $k, k_{1}, k_{2}, k_{4}, \lambda$ at $1-\sigma, 2-\sigma$ and $3-\sigma$ confidence level in MCMC analysis of $H(z)$ datasets.}
	\end{figure}
	%%%%%%%%%%%%%%%%%%%%%%%%%%%%%%%%%%%%%%%%%%%%%%%%%%%%%%%%%%%%%%%%%%
	%%%%%%%%%%%%%%%%%%%%%%%%%%%%%%%%%%%%%%%%%%%%%%%%%%%%%%%%%%%%
	%%%%%%%%%%%%%%%%%%%%%%%%%%%%%%%%%%%% Figure 3
	%%%%%%%%%%%%%%%%%%%%%%%%%%%%%%%%%%%%%%%%%%%%%%%%%%%%%%%%%%%%
	\begin{figure}[H]
		\centering
		a.\includegraphics[width=8cm,height=6cm,angle=0]{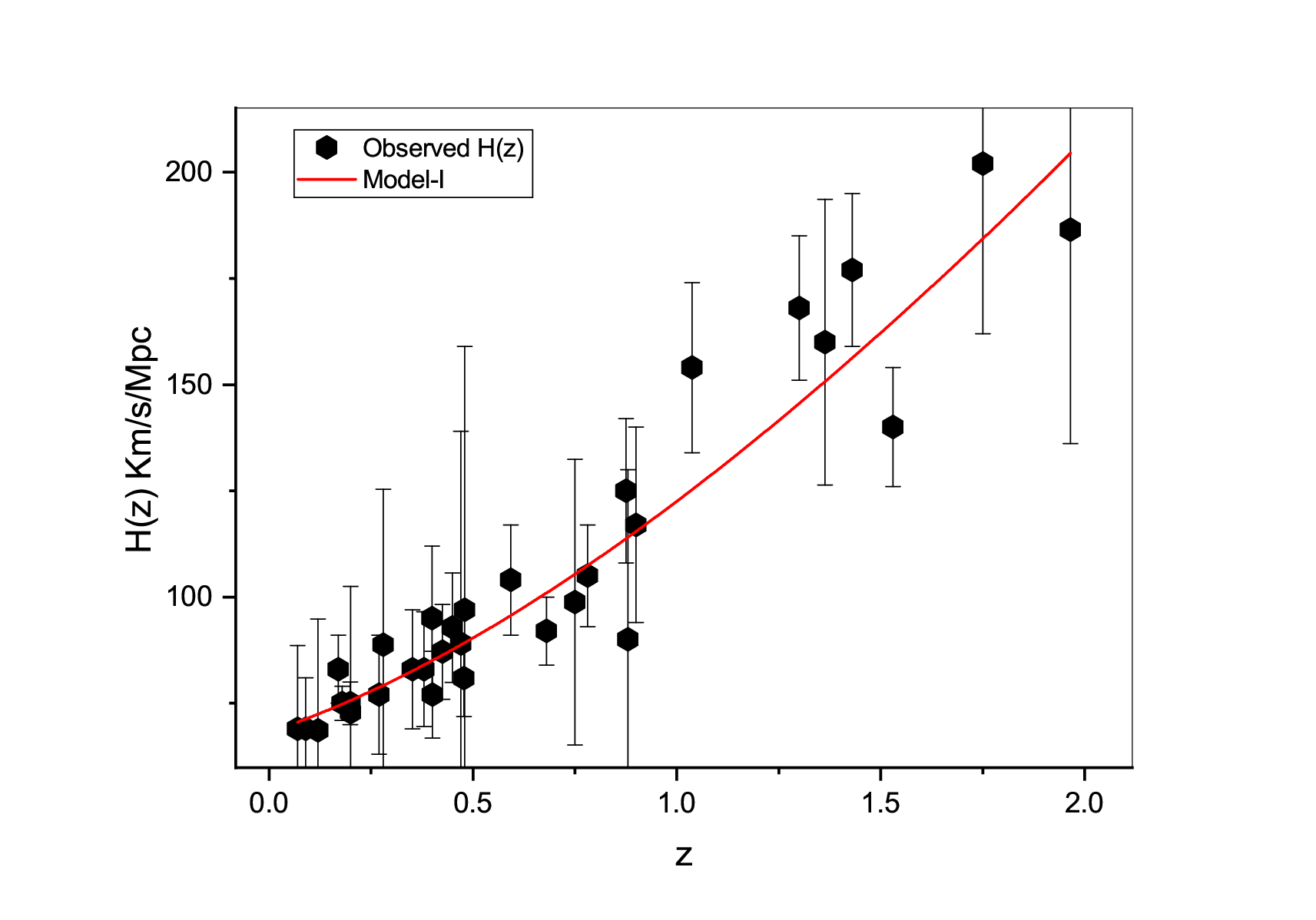}
		b.\includegraphics[width=8cm,height=6cm,angle=0]{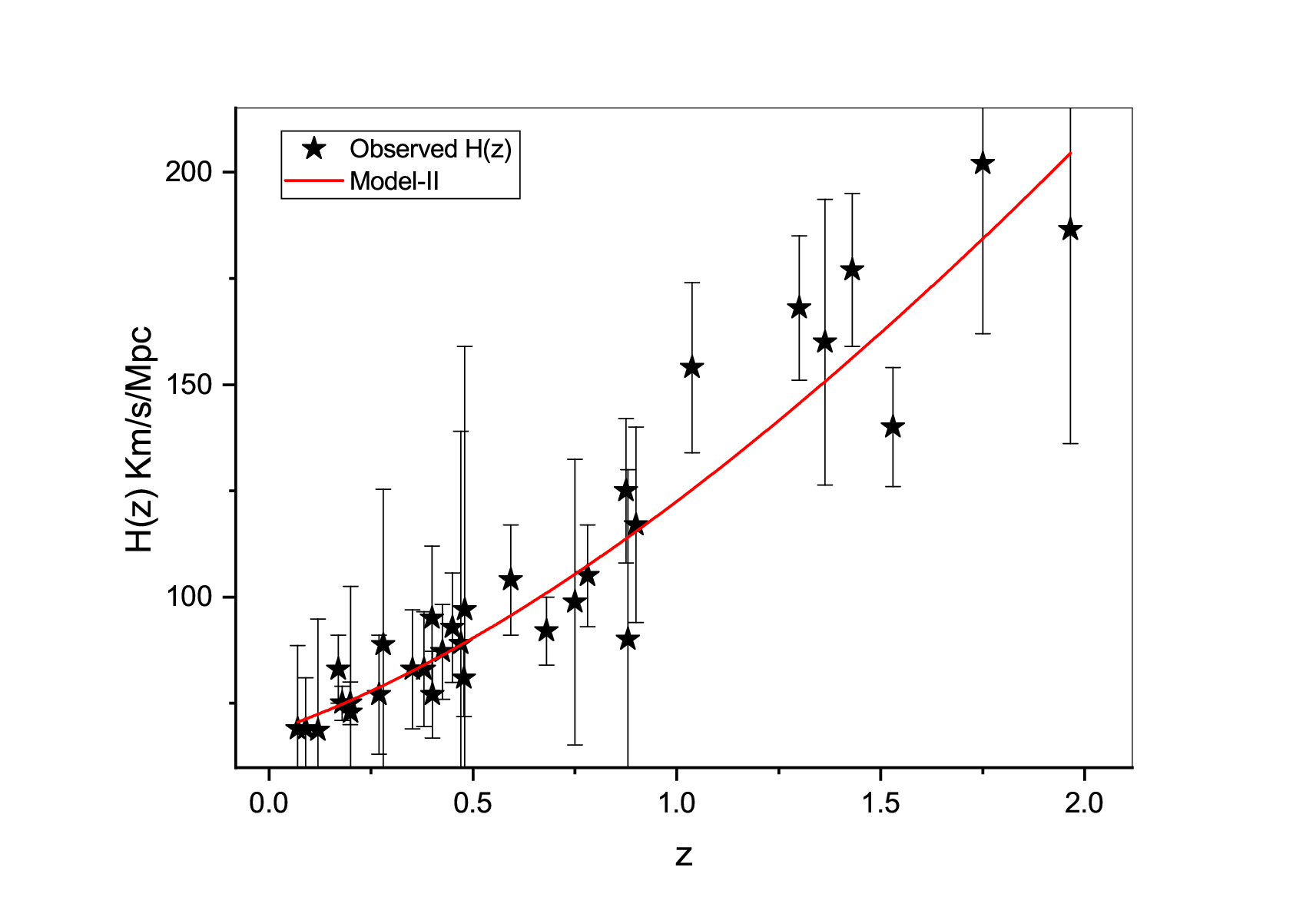}
		\caption{The Hubble error bar plots of $H(z)$ datasets over $z$, respectively for Model-I, and Model-II.}
	\end{figure}
	%%%%%%%%%%%%%%%%%%%%%%%%%%%%%%%%%%%%%%%%%%%%%%%%%%%%%%%%%%%%%%%%%%
	Figure 1 \& 2 represent the contour plots of model parameters involved in Model-I \& Model-II, respectively, with observed Hubble datasets $H(z)$, using MCMC analysis. We have obtained the best fit values of various model parameters $k, k_{1}, k_{2}, k_{4}, \lambda$, by using the different values of parameters in the range $50<k<100$, $-1<k_{1}<2.5$, $10000<k_{2}<20000$, $10000<k_{4}<30000$ and $-0.7<\lambda<3.0$ with $1-\sigma, 2-\sigma$ and $3-\sigma$ confidence level. The best fit values are mentioned in Table 1, for both models, respectively. We have estimated the best fit values of model parameter $\lambda=0.34563_{-0.25994}^{+0.27994}, 0.20855_{-0.15423}^{+0.21128}$ for Model-I, Model-II, respectively. Recently \cite{ref55} has estimated the value of $\lambda$ as $\lambda=0.491_{-0.533}^{+0.387}, 0.537_{-0.550}^{+0.403}$, respectively, in two different models. Figure 3a \& 3b depicts the Hubble error bar plots for Model-I \& Model-II, respectively.
	%%%%%%%%%%%%%%%%%%%%%%%%%%%%%%%%%%%%%%%%%%%%%%%%%%%%%%%%%%%%%%%%%
	\subsection{Apparent Magnitude $m(z)$}
	%%%%%%%%%%%%%%%%%%%%%%%%%%%%%%%%%%%%%%%%%%%%%%%%%%%%%%%%%%%%%%%%%
	    The relationship between luminosity distance and redshift is one of the main observational techniques used to track the universe's evolution. The expansion of the cosmos and the redshift of the light from distant brilliant objects are taken into consideration when calculating the luminosity distance ($D_{L}$) in terms of the cosmic redshift ($z$). It is provided as
	\begin{equation}\label{eq45}
		D_{L}=a_{0} r (1+z),
	\end{equation}
	where the radial coordinate of the source $r$, is established by
	\begin{equation}\label{eq46}
		r  =  \int^r_{0}dr = \int^t_{0}\frac{cdt}{a(t)} = \frac{1}{a_{0}}\int^z_0\frac{cdz}{H(z)},
	\end{equation}
	where we have used $ dt=dz/\dot{z}, \dot{z}=-H(1+z)$.\\
	As a result, the luminosity distance is calculated as follows:
	\begin{equation}\label{eq47}
		D_{L}=c(1+z)\int^z_0\frac{dz}{H(z)}.
	\end{equation}
	Hence, the apparent magnitude $m(z)$ of a supernova is defined as:
	\begin{equation}\label{eq48}
		m(z)=16.08+ 5~log_{10}\left[\frac{(1+z)H_{0}}{0.026} \int^z_0\frac{dz}{H(z)}\right].
	\end{equation}
	We use the most recent collection of $1048$ datasets of the Pantheon SNe Ia samples in the ($0.01 \le z \le 1.7$) range \cite{ref79} in our MCMC analysis. We have used the following $\chi^{2}$ formula to constrain different model parameters:
	\begin{equation}\nonumber
		\chi^{2}(\phi)=\sum_{i=1}^{i=N}\frac{[(m_{ob})_{i}-(m_{th})_{i}]^{2}}{\sigma_{i}^{2}}.
	\end{equation}
	The entire amount of data is denoted by $N$, the observed and theoretical datasets of $m(z)$ are represented by $m_{ob}$ and $m_{th}$, respectively, and standard deviations are denoted by $\sigma_{i}$. Here for the Model-I $\phi=(H_{0}, k, k_{2}, k_{4}, \lambda)$ and for the Model-II $\phi=(H_{0}, k, k_{1}, k_{2}, k_{4}, \lambda)$.
	%%%%%%%%%%%%%%%%%%%%%%%%%%%%%%%%%%%%%%%%%%%%%%%%%%%%%%%%%%%%
	%%%%%%%%%%%%%%%%%%%%%%%%%%%%%%%%%%%%% Figure 4
	%%%%%%%%%%%%%%%%%%%%%%%%%%%%%%%%%%%%%%%%%%%%%%%%%%%%%%%%%%%%
	\begin{figure}[H]
		\centering
		\includegraphics[width=9cm,height=9cm,angle=0]{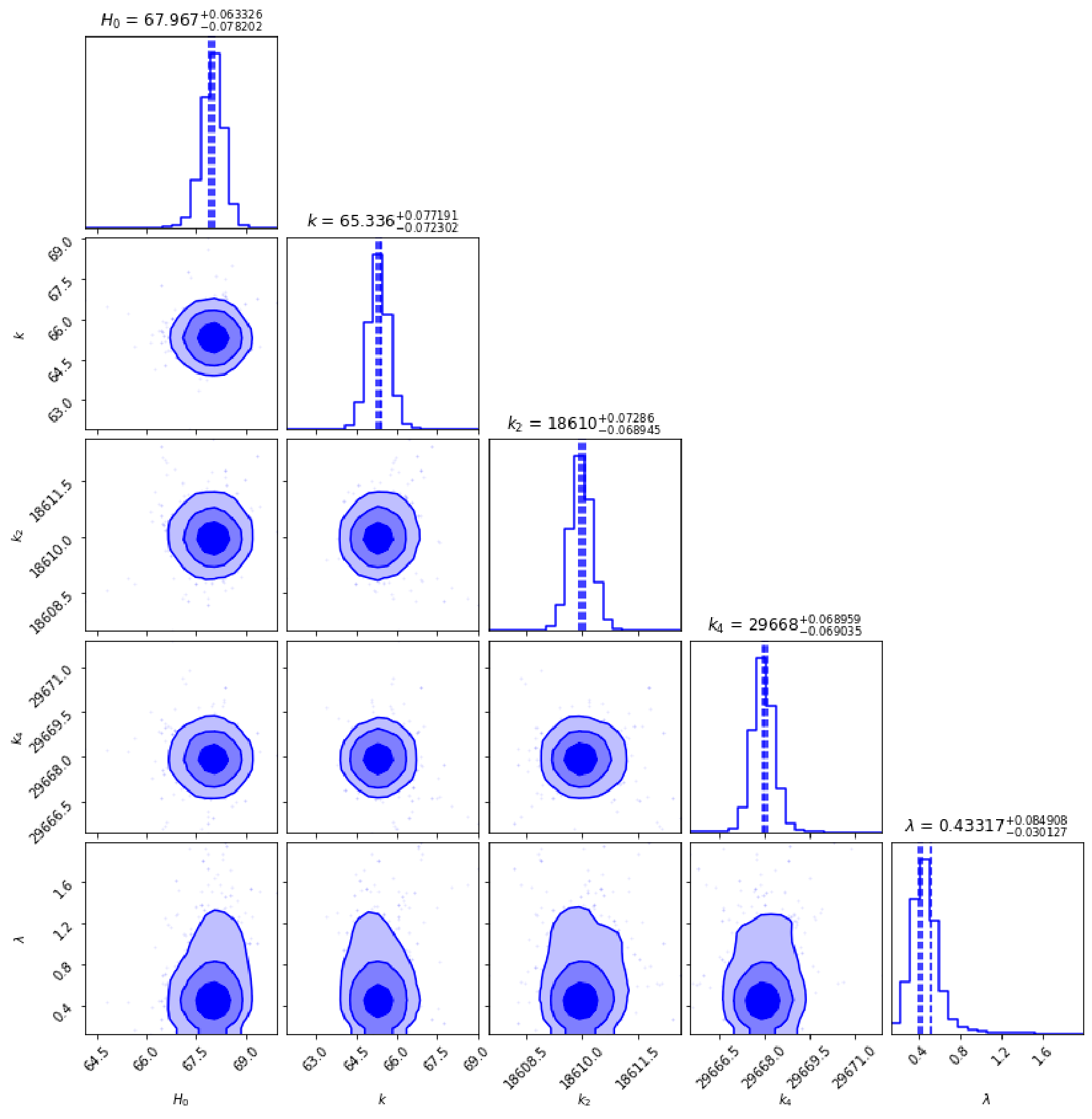}
		\caption{The contour plots of $H_{0}, k, k_{2}, k_{4}, \lambda$ at $1-\sigma, 2-\sigma$ and $3-\sigma$ confidence level in MCMC analysis of Pantheon SNe Ia datasets.}
	\end{figure}
	%%%%%%%%%%%%%%%%%%%%%%%%%%%%%%%%%%%%%%%%%%%%%%%%%%%%%%%%%%%%%%%%%%
		%%%%%%%%%%%%%%%%%%%%%%%%%%%%%%%%%%%%%%%%%%%%%%%%%%%%%%%%%%%%
	%%%%%%%%%%%%%%%%%%%%%%%%%%%%%%%%%%%%% Figure 5
	%%%%%%%%%%%%%%%%%%%%%%%%%%%%%%%%%%%%%%%%%%%%%%%%%%%%%%%%%%%%
	\begin{figure}[H]
		\centering
		\includegraphics[width=11cm,height=11cm,angle=0]{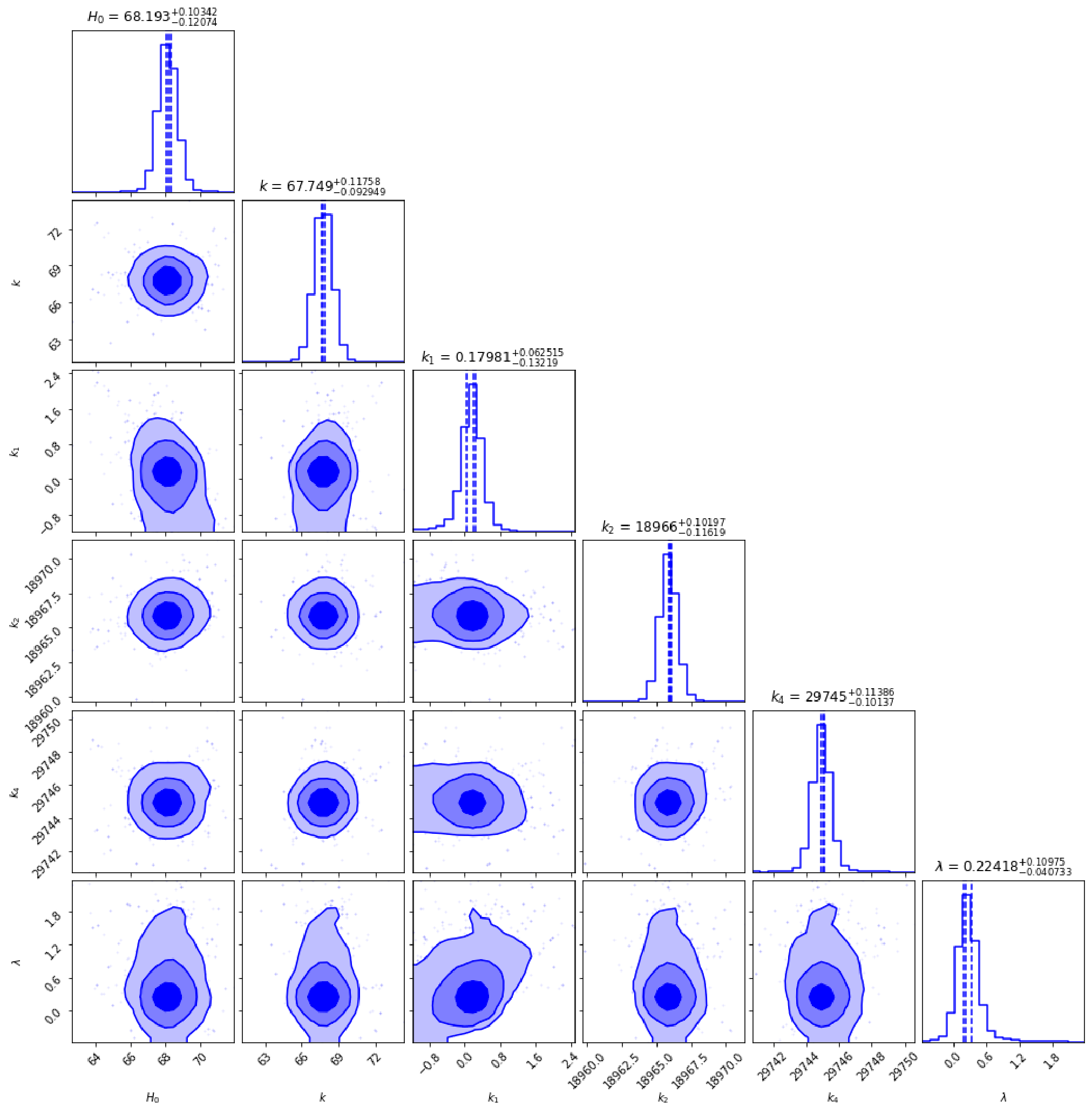}
		\caption{The contour plots of $H_{0}, k, k_{1}, k_{2}, k_{4}, \lambda$ at $1-\sigma, 2-\sigma$ and $3-\sigma$ confidence level in MCMC analysis of Pantheon SNe Ia datasets.}
	\end{figure}
	%%%%%%%%%%%%%%%%%%%%%%%%%%%%%%%%%%%%%%%%%%%%%%%%%%%%%%%%%%%%%%%%%%
			%%%%%%%%%%%%%%%%%%%%%%%%%%%%%%%%%%%%%%%%%%%%%%%%%%%%%%%%%%%%%%%%%%%%%	
	\begin{table}[H]
		\centering
		\begin{tabular}{|c|c|c|c|}
			\hline
			% after \\: \hline or \cline{col1-col2} \cline{col3-col4} ...
			
			Parameter       & Prior                    & Model-I  & Model-II   \\
			\hline
			$H_{0}$         & $(50, 100)$              & $67.967_{-0.078202}^{+0.063326}$  & $68.193_{-0.12074}^{+0.10342}$ \\
			$k$             & $(50, 100)$              & $65.336_{-0.072302}^{+0.077101}$  & $67.749_{-0.092949}^{+0.11758}$ \\
			$k_{1}$         & $(-1.0, 2.5)$            & $-$                               & $0.17981_{-0.13219}^{+0.062515}$ \\
			$k_{2}$         & $(10^{4}, 2\times10^{4})$& $18610_{-0.068945}^{+0.07286}$    & $18966_{-0.11619}^{+0.10197}$ \\
			$k_{4}$         & $(10^{4}, 3\times10^{4})$& $29668_{-0.069035}^{+0.068959}$   & $29745_{-0.10137}^{+0.11386}$ \\
			$\lambda$       & $(-0.7, 3.0)$            & $0.43317_{-0.030127}^{+0.084908}$ & $0.22418_{-0.040733}^{+0.10975}$ \\
			\hline
		\end{tabular}
		\caption{The MCMC Results in Pantheon SNe Ia datasets analysis.}\label{T2}
	\end{table}
	%%%%%%%%%%%%%%%%%%%%%%%%%%%%%%%%%%%%%%%%%%%%%%%%%%%%%%%%%%%%%%%%%%%%%%%%%%%%
		Figure 4 \& 5 represent the contour plots of model parameters involved in Model-I \& Model-II, respectively, with apparent magnitude $m(z)$ from observed Pantheon SNe Ia datasets, using MCMC analysis. We have obtained the best fit values of various model parameters $H_{0}, k, k_{1}, k_{2}, k_{4}, \lambda$, by using the different values of parameters in the range $40<H_{0}<100$, $50<k<100$, $-1<k_{1}<2.5$, $10000<k_{2}<20000$, $10000<k_{4}<30000$ and $-0.7<\lambda<3.0$ with $1-\sigma, 2-\sigma$ and $3-\sigma$ confidence level. The best fit values of various model parameters are mentioned in Table 2, for both models, respectively. We have estimated the best fit values of model parameter $\lambda=0.43317_{-0.030127}^{+0.084908}, 0.22418_{-0.040733}^{+0.10975}$ for Model-I, Model-II, respectively. Recently \cite{ref55} has estimated the value of $\lambda$ as $\lambda=0.491_{-0.533}^{+0.387}, 0.537_{-0.550}^{+0.403}$, respectively, in two different models. We have estimated the value of Hubble constant as $H_{0}=67.967_{-0.078202}^{+0.063326}, 68.193_{-0.12074}^{+0.10342}$ Km/s/Mpc, respectively for two models, Model-I \& Model-II along Pantheon SNe Ia datasets. Recently, the value of Hubble constant is measured as $H_{0}=69.8\pm1.3~Km s^{-1} Mpc^{-1}$ in \cite{ref80}, in \cite{ref81}, it is measured as $H_{0}=69.7\pm1.2~Km s^{-1} Mpc^{-1}$. Recently, we have measured the value of $H_{0}$ as $H_{0}=68.3721\pm1.7205$ Km/s/Mpc in \cite{ref59}, $H_{0}=68.3721\pm1.65678$ Km/s/Mpc in \cite{ref57} and $H_{0}=71.66123\pm0.33061$ in \cite{ref58}.		
		
	%%%%%%%%%%%%%%%%%%%%%%%%%%%%%%%%%%%%%%%%%%%%%%%%%%%%%%%%%%%%%%%%%
	\section{Result Discussion}
	%%%%%%%%%%%%%%%%%%%%%%%%%%%%%%%%%%%%%%%%%%%%%%%%%%%%%%%%%%%%%%%%%
	First, we discuss each model in details as follows:
	%%%%%%%%%%%%%%%%%%%%%%%%%%%%%%%%%%%%%%%%%%%%%%%%%%%%%%%%%%%%%%%%
	\subsection{Model-I}
	%%%%%%%%%%%%%%%%%%%%%%%%%%%%%%%%%%%%%%%%%%%%%%%%%%%%%%%%%%%%%%%%
	The scale factor $a(t)$ is the most powerful parameter in cosmology that governs every geometrical parameters behaviour. From the solution of the field equation \eqref{eq16}, we obtain the scale factor $a(t)$ as in \eqref{eq19}, and its geometrical evolution over cosmic time $t$ is given in figure 6a. From the figure 6a, we can observe that the scale factor $a(t)$ is an increasing function of cosmic time $t$ that shows that our universe model is expanding. One can see that as $t\to t_{0}$ (present time) then the value of $a(t)\to1$ (i.e. $a_{0}=a(t_{0})=1$), and as $t\to\infty$ then $a(t)\to\infty$. The present age of the universe is estimated as $t_{0}=0.01363_{-0.00020}^{+0.00027}, 0.01379_{-0.00006}^{+0.00003}$ or $t_{0}=13.33_{-0.19}^{+0.26}, 13.48_{-0.05}^{+0.02}$ Gyrs, respectively along two datasets $H(z)$ and Pantheon SNe Ia. In general, the present age of the universe can be expressed as $t_{0}=978\times\frac{2}{\sqrt{3}}\sqrt{\frac{2(1+\lambda)}{k_{2}+\lambda k_{4}}}\sinh^{-1}\left(k\sqrt{\frac{2(1+\lambda)}{k_{2}+\lambda k_{4}}}\right)$ Gyrs.
	%%%%%%%%%%%%%%%%%%%%%%%%%%%%%%%%%%%%%%%%%%%%%%%%%%%%%%%%%%%%
	%%%%%%%%%%%%%%%%%%%%%%%%%%%%%%%%%%%% Figure 6
	%%%%%%%%%%%%%%%%%%%%%%%%%%%%%%%%%%%%%%%%%%%%%%%%%%%%%%%%%%%%
	\begin{figure}[H]
		\centering
		a.\includegraphics[width=8cm,height=6cm,angle=0]{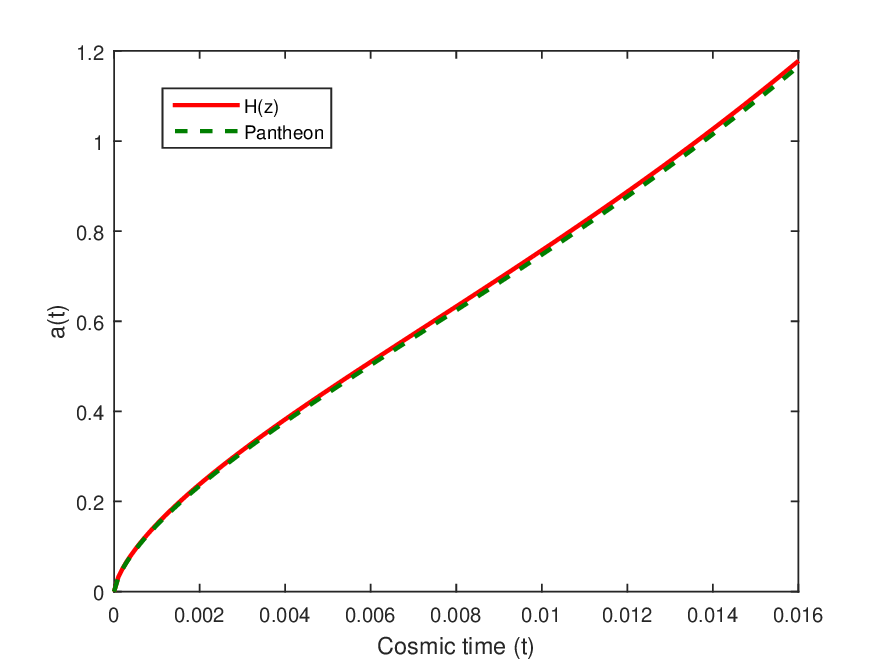}
		b.\includegraphics[width=8cm,height=6cm,angle=0]{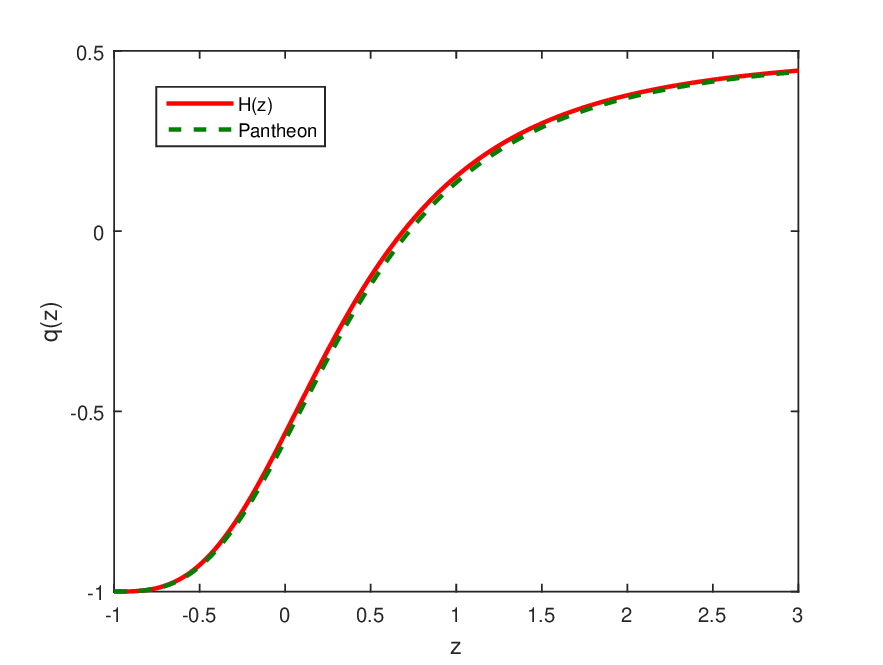}
		\caption{Variation of scale factor over cosmic time $t$, and deceleration parameter $q(z)$ over $z$, respectively.}
	\end{figure}
	%%%%%%%%%%%%%%%%%%%%%%%%%%%%%%%%%%%%%%%%%%%%%%%%%%%%%%%%%%%%%%%%%%
	The geometrical parameter $q(t)$ is defined in terms of scale factor as $q(t)=-\frac{a\ddot{a}}{\dot{a}^{2}}$ that reveals the expansion phase of the universe (decelerating or accelerating). In our model-I, it is derived as in Eq.~\eqref{eq21} from the scale factor \eqref{eq19}, and expressed in terms of redshift $z$ as in Eq.~\eqref{eq23}. The geometrical behaviour of deceleration parameter $q(z)$ is shown in figure 6b, and one can see that $q(z)$ is an increasing function of redshift $z$ and shows a signature-flipping point (decelerating-accelerating) called as transition redshift $z$. The present value of $q(z)$ is obtained as $q_{0}=-0.5610_{-0.0141}^{+0.0254}, -0.58_{-0.0051}^{+0.0016}$, respectively for two observational datasets $H(z)$ and Pantheon SNe Ia, that reveals that our universe model is accelerating phase of expansion at present (at $z=0$). We observe that as $z\to-1$, then $q\to-1$ (late-time acceleration) while as $z\to\infty$ then $q\to0.5$ early decelerating universe. This model can not explain the early accelerating (inflation) scenario of the evolution of the universe. Also, one can obtain as $z\to0$, then the value of deceleration parameter $q\to-1+\frac{3}{2}\left[1+\frac{k_{2}+\lambda k_{4}}{2k^{2}(1+\lambda)}\right]^{-1}$. The estimated transition redshift is obtained as $z_{t}=0.69_{-0.056}^{+0.028}, 0.726_{-0.004}^{+0.009}$, respectively along two datasets, and the general expression for transition redshift is obtained as
	\begin{equation}\label{eq49}
		z_{t}=\left[\frac{k_{2}+\lambda k_{4}}{k^{2}(1+\lambda)}\right]^{\frac{1}{3}}-1,~~~~\lambda\ne-1.
	\end{equation}
	This transition value shows that the universe model is in decelerating phase of expansion for $z>z_{t}$ and the model is in accelerating phase of expansion for $z<z_{t}$. Recently, this transition redshift $z_{t}=0.74\pm0.05$ is obtained in \cite{ref82} and in \cite{ref83} it is measured as $z_{t}=0.74\pm0.04$. In \cite{ref84}, this transition redshift is obtained as $z_{t}=0.72\pm0.05$ and in 2018, \cite{ref85} has suggested that the transition redshift varies over $0.33 < z_{t} < 1.0$. Recently, we have found this transition redshift $z_{t}\approx0.7$ in \cite{ref86,ref87}. Thus, the transition redshift $z_{t}=0.69_{-0.056}^{+0.028}, 0.726_{-0.004}^{+0.009}$ obtained in our derived model is in good agreement with recent observed values in \cite{ref82}-\cite{ref87}.\\
	
	The cosmological parameters $\Omega_{m}$ and $\Omega_{F}$ are derived from the field Eqs.~\eqref{eq15} as in Eqs.~\eqref{eq29} \& \eqref{eq30} for the model-I. Here, $\Omega_{m}$ is the non-relativistic matter energy density parameter and $\Omega_{F}$ is the dark energy density parameter from geometrical modifications. The geometrical behaviour of $\Omega_{m}$ \& $\Omega_{F}$ are shown in figure 7a \& 7b, respectively. The estimated present values of total energy density parameters are $\Omega_{m0}=0.3938_{-0.0577}^{+0.0656}, 0.4013_{-0.0069}^{+0.0186}$ and $\Omega_{F0}=0.6062_{-0.0656}^{+0.0577}, 0.5987_{-0.0186}^{+0.0069}$, respectively along two observational datasets $H(z)$ and Pantheon SNe Ia. From figure 7a \& 7b, one can see that $\Omega_{m}\to0$ \& $\Omega_{F}\to1$ as $z\to-1$ (at late-time) which reveals that our model tends to $\Lambda$CDM model at late-time universe, and in early universe $\Omega_{m}\to1$ and $\Omega_{F}\to-\lambda$ as $z\to\infty$ that shows the matter dominated early universe.
 	%%%%%%%%%%%%%%%%%%%%%%%%%%%%%%%%%%%%%%%%%%%%%%%%%%%%%%%%%%%%
	%%%%%%%%%%%%%%%%%%%%%%%%%%%%%%%%%%%% Figure 7
	%%%%%%%%%%%%%%%%%%%%%%%%%%%%%%%%%%%%%%%%%%%%%%%%%%%%%%%%%%%%
	\begin{figure}[H]
		\centering
		a.\includegraphics[width=8cm,height=6cm,angle=0]{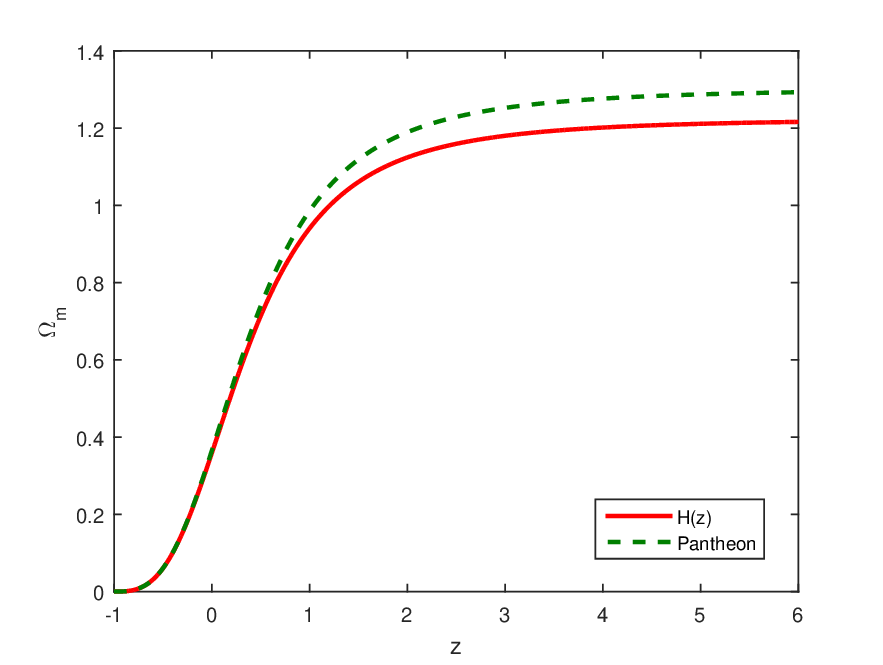}
		b.\includegraphics[width=8cm,height=6cm,angle=0]{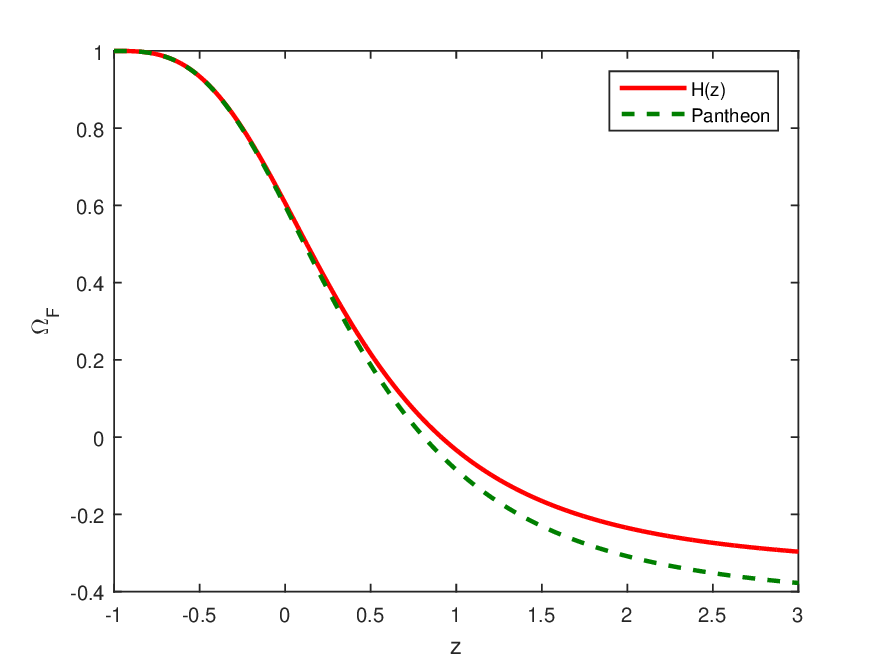}
		\caption{The plot of total energy density parameters $\Omega_{m}$ and $\Omega_{F}$ over redshift $z$, respectively.}
	\end{figure}
	%%%%%%%%%%%%%%%%%%%%%%%%%%%%%%%%%%%%%%%%%%%%%%%%%%%%%%%%%%%%%%%%%%
	%%%%%%%%%%%%%%%%%%%%%%%%%%%%%%%%%%%%%%%%%%%%%%%%%%%%%%%%%%%%
	%%%%%%%%%%%%%%%%%%%%%%%%%%%%%%%%%%%% Figure 8
	%%%%%%%%%%%%%%%%%%%%%%%%%%%%%%%%%%%%%%%%%%%%%%%%%%%%%%%%%%%%
	\begin{figure}[H]
		\centering
		a.\includegraphics[width=8cm,height=6cm,angle=0]{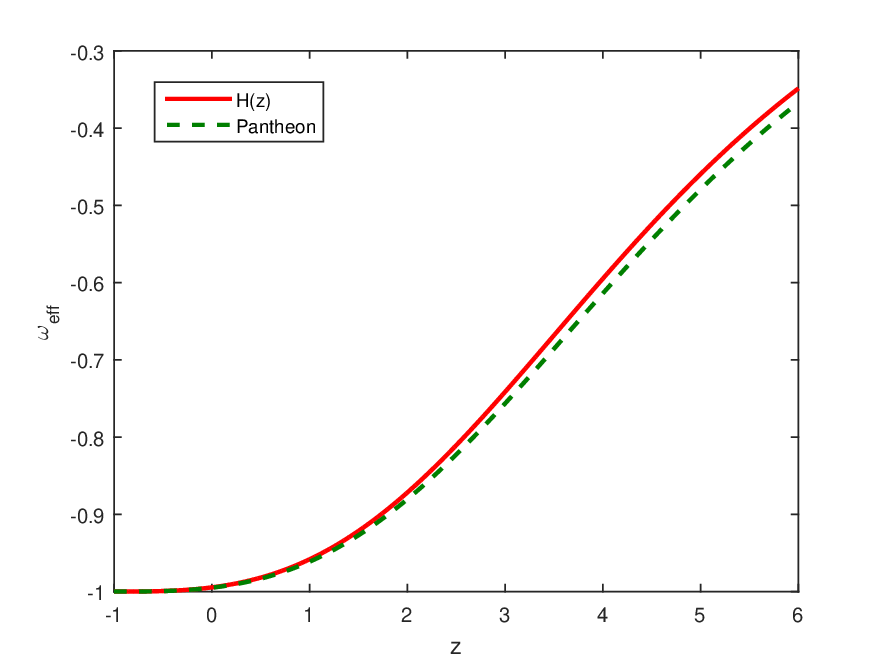}
		b.\includegraphics[width=8cm,height=6cm,angle=0]{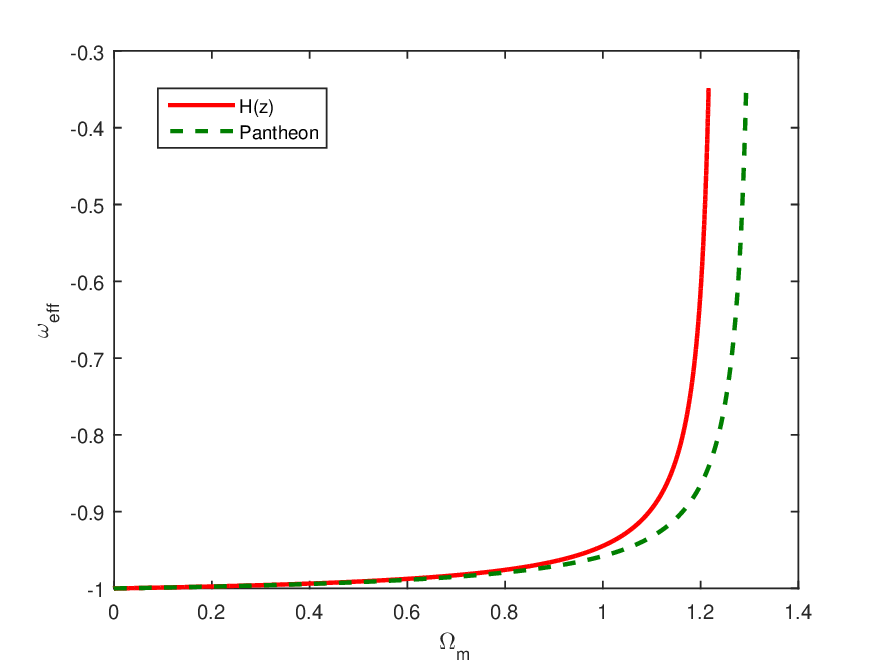}
		\caption{The plot of effective EoS parameter $\omega_{eff}$ versus $z$ \& $\Omega_{m}$, respectively.}
	\end{figure}
	%%%%%%%%%%%%%%%%%%%%%%%%%%%%%%%%%%%%%%%%%%%%%%%%%%%%%%%%%%%%%%%%%%
	The effective EoS parameter $\omega_{eff}$ for model-I is derived in Eq.~\eqref{eq28}, and its geometrical evolution over $z$ is shown in figure 8a. From the figure 8a, one can see that $\omega_{eff}\to-1$ as $z\to-1$ (at late-time) and $\omega_{eff}\to0$ as $z\to\infty$ (at early universe time), that support early matter dominated universe and late-time dark energy dominated universe. The present value of effective EoS parameter for model-I is estimated as $\omega_{eff}=-0.9946_{-0.0020}^{+0.0024}, -0.9950_{-0.0006}^{+0.0002}$, respectively along two observational datasets. Figure 8b reveals the variation of $\omega_{eff}$ over matter energy density parameter $\Omega_{m}$ and mathematically we can find this relationship as $(A-B)k^{2}\Omega_{m}\omega_{eff}+B\rho_{m0}\kappa^{2}\omega_{eff}-Bk^{2}\Omega_{m}+B\rho_{m0}\kappa^{2}=0$, where $A=2n_{1}(\kappa^{2}\rho_{m0}-\lambda k^{2})$ and $B=(k_{2}+\lambda k_{4}-2n_{1}\lambda)k^{2}$. Also, it can be expressed as
	\begin{equation}\label{eq50}
		\omega_{eff}=\frac{Bk^{2}\Omega_{m}-B\rho_{m0}\kappa^{2}}{(A-B)k^{2}\Omega_{m}+B\rho_{m0}\kappa^{2}}
	\end{equation}
	From Eq.~\eqref{eq50} and figure 7b, it is clear that the EoS parameter $\omega_{eff}\to-1$ for $\Omega_{m}\to0$, and $\omega_{eff}\to0$ for $\Omega_{m}\to\frac{\rho_{m0}\kappa^{2}}{k^{2}}$. From the above relationship, it is predicted that one can obtain radiation dominated universe for $\Omega_{m}>\frac{\rho_{m0}\kappa^{2}}{k^{2}}$.\\
	
	Now, we discuss two more geometrical parameters derived from the scale factor $a(t)$ as the statefinder diagnostic parameters $r(t)$, $s(t)$ which are defined by the Eq.~\eqref{eq24}. For the model-I, we have derived these parameters $r(t)$ \& $s(t)$ as in Eqs.~\eqref{eq25} \& \eqref{eq26}, respectively. The geometrical evolution of these parameters $r, s$ are shown in figure 9a \& 9b, over redshift $z$, respectively. From the figure 9a, 9b, one can see that $r(z)\to1$, $s(z)\to0$ as $z\to-1$, that reveals the late-time $\Lambda$CDM tendency of the model. The estimated present values of $r, s$ are as $r_{0}=0.122_{-0.0508}^{+0.0302}, 0.16_{-0.0033}^{+0.0103}$ and $s_{0}=0.2759_{-0.0133}^{+0.0231}, 0.2593_{-0.0044}^{+0.0014}$, respectively along two datasets $H(z)$ and Pantheon SNe Ia.
	%%%%%%%%%%%%%%%%%%%%%%%%%%%%%%%%%%%%%%%%%%%%%%%%%%%%%%%%%%%%
	%%%%%%%%%%%%%%%%%%%%%%%%%%%%%%%%%%%% Figure 9
	%%%%%%%%%%%%%%%%%%%%%%%%%%%%%%%%%%%%%%%%%%%%%%%%%%%%%%%%%%%%
	\begin{figure}[H]
		\centering
		a.\includegraphics[width=8cm,height=6cm,angle=0]{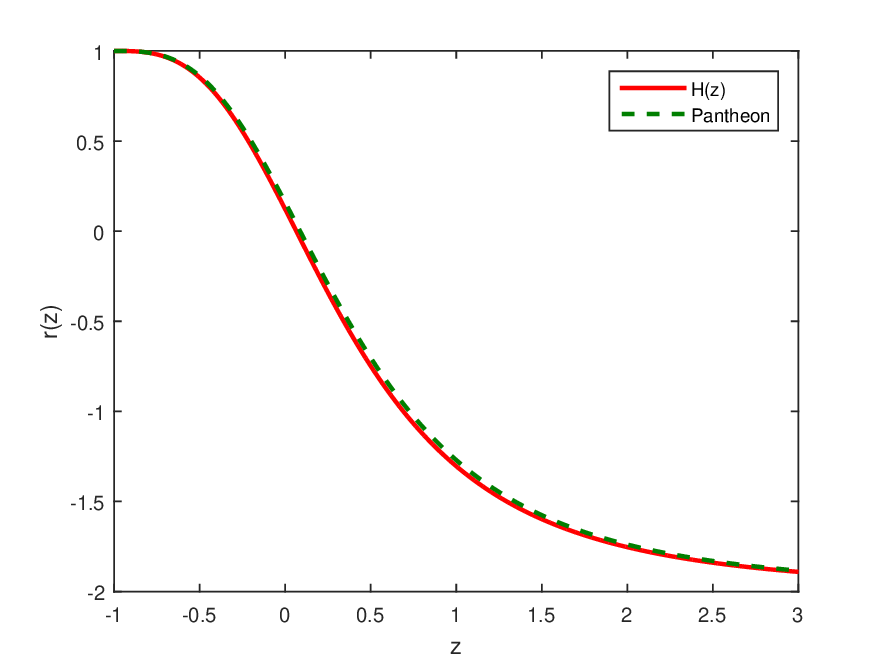}
		b.\includegraphics[width=8cm,height=6cm,angle=0]{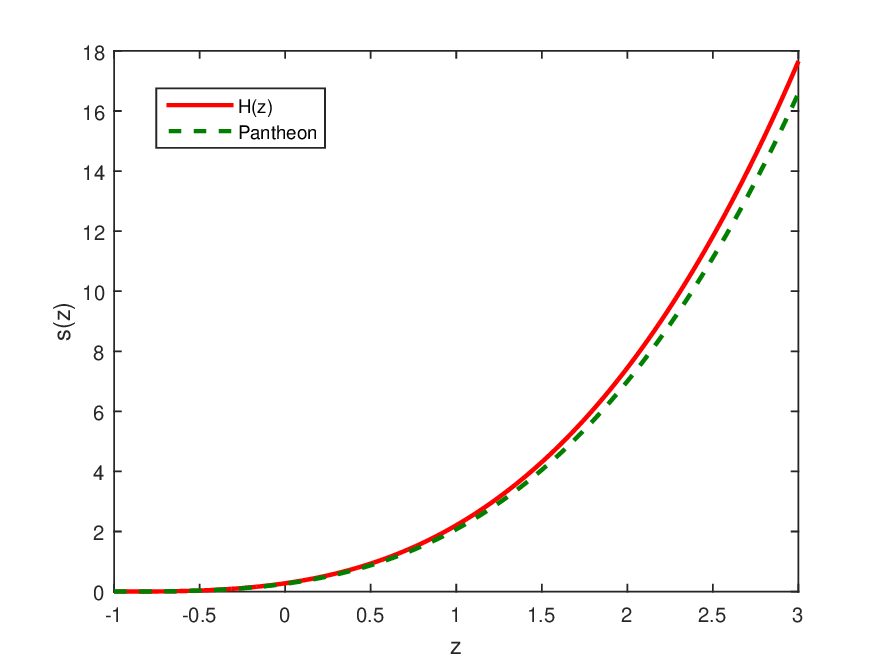}
		\caption{The plot of statefinder diagnostic parameter $r(z)$ \& $s(z)$ versus redshift $z$, respectively.}
	\end{figure}
	%%%%%%%%%%%%%%%%%%%%%%%%%%%%%%%%%%%%%%%%%%%%%%%%%%%%%%%%%%%%%%%%%%
	%%%%%%%%%%%%%%%%%%%%%%%%%%%%%%%%%%%%%%%%%%%%%%%%%%%%%%%%%%%%
	%%%%%%%%%%%%%%%%%%%%%%%%%%%%%%%%%%%% Figure 10
	%%%%%%%%%%%%%%%%%%%%%%%%%%%%%%%%%%%%%%%%%%%%%%%%%%%%%%%%%%%%
	\begin{figure}[H]
		\centering
		a.\includegraphics[width=8cm,height=6cm,angle=0]{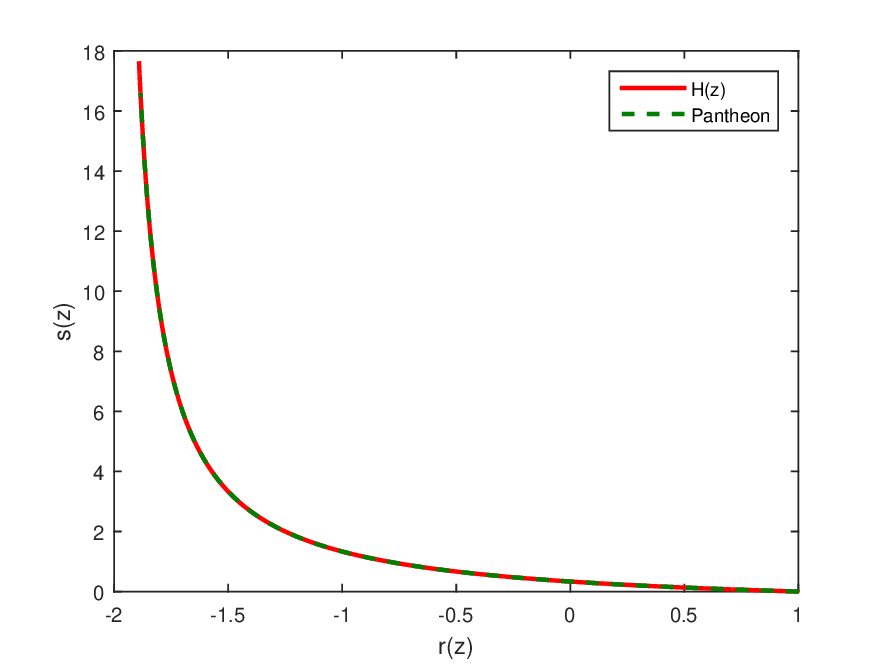}
		b.\includegraphics[width=8cm,height=6cm,angle=0]{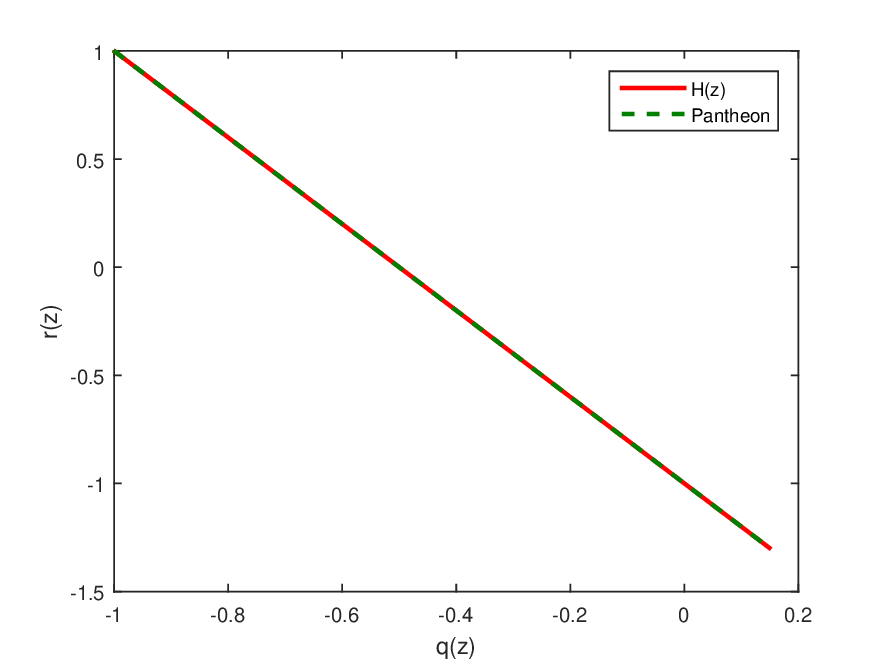}
		\caption{The variation of statefinder diagnostic parameter $s(z)$ versus $r(z)$, and $r(z)$ versus $q(z)$, respectively.}
	\end{figure}
	%%%%%%%%%%%%%%%%%%%%%%%%%%%%%%%%%%%%%%%%%%%%%%%%%%%%%%%%%%%%%%%%%%
	Figure 10a represents the variation of statefinder parameter $s(z)$ with the variation of statefinder parameter $r(z)$ that reveals that these parameters are related to each other. Also, we can find the relation between $r$ \& $s$ from Eqs.~\eqref{eq25} \& \eqref{eq26} as $3rs+6s+2r-2=0$ for the model-I. It can be expressed as $s=\frac{2(1-r)}{3(2+r)}$ which shows that $s\to\infty$ for $r\to-2$ (singular point) and $s\to0$ for $r\to1$. Figure 10b shows the relationship between statefinder parameter $r$ and deceleration parameter $q$ that reveals that $r$ \& $q$ are linearly related to each other. From the Eqs.~\eqref{eq21} \& \eqref{eq25}, we can find the relationship between $r$ \& $q$ as $r+2q+1=0$ for the model-I. Also, we can rewrite as $q=-\frac{1+r}{2}$ which gives $q>0$ for $r<-1$ and $q<0$ for $r>-1$, and the model undergoes a transition point at $r=-1$.
	%%%%%%%%%%%%%%%%%%%%%%%%%%%%%%%%%%%%%%%%%%%%%%%%%%%%%%%%%%%%
	%%%%%%%%%%%%%%%%%%%%%%%%%%%%%%%%%%%% Figure 11
	%%%%%%%%%%%%%%%%%%%%%%%%%%%%%%%%%%%%%%%%%%%%%%%%%%%%%%%%%%%%
	\begin{figure}[H]
		\centering
		a.\includegraphics[width=8cm,height=6cm,angle=0]{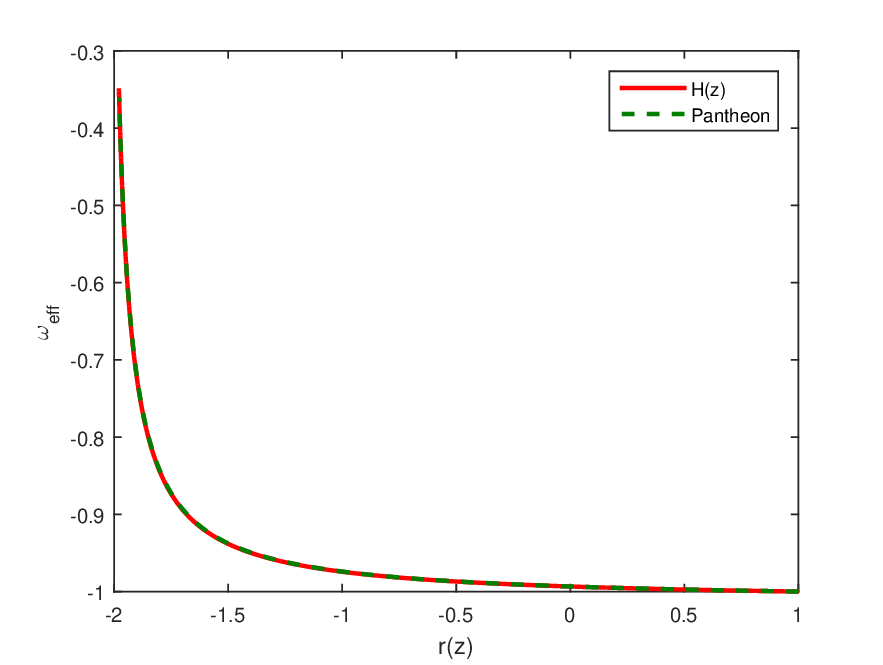}
		b.\includegraphics[width=8cm,height=6cm,angle=0]{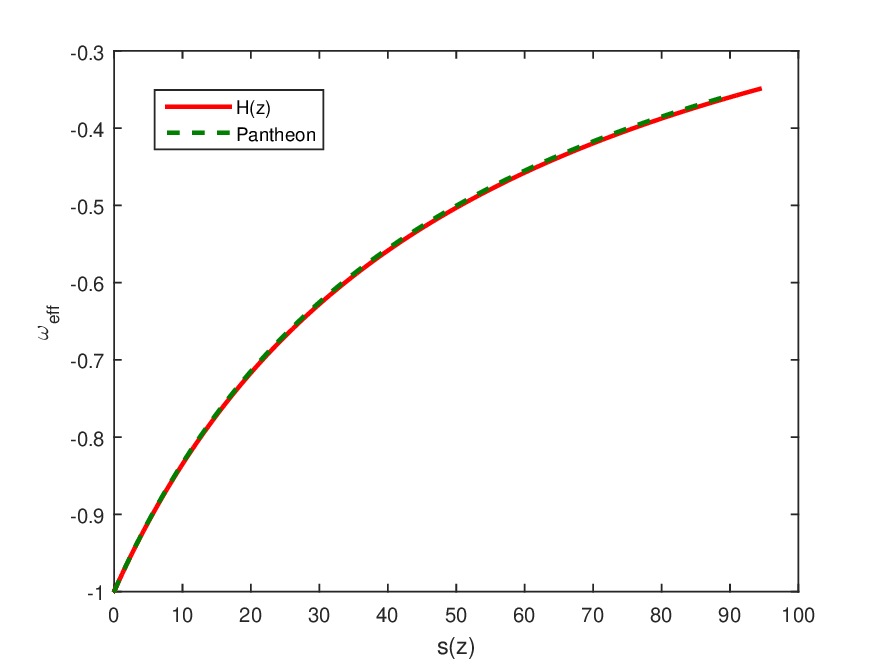}
		\caption{The plots of effective EoS parameter $\omega_{eff}$ versus statefinder parameters $r(z)$ and $s(z)$, respectively.}
	\end{figure}
	%%%%%%%%%%%%%%%%%%%%%%%%%%%%%%%%%%%%%%%%%%%%%%%%%%%%%%%%%%%%%%%%%%
	Figure 11a depicts the evolution of effective EoS parameter $\omega_{eff}$ over the variation of statefinder parameter $r(z)$, that depicts $\omega_{eff}$ \& $r$ are related each other. From the Eqs.~\eqref{eq25} \& \eqref{eq28}, one can obtain the following relationship between $\omega_{eff}$ \& $r$ as $(B-A)r\omega_{eff}+(A+2B)\omega_{eff}+Br+2B=0$ for the model-I, where $A=2n_{1}(\kappa^{2}\rho_{m0}-\lambda k^{2})$ and $B=(k_{2}+\lambda k_{4}-2n_{1}\lambda)k^{2}$. It can be expressed as $\omega_{eff}=\frac{B(r+2)}{A(r-1)-B(r+2)}$ that shows as $r\to1$ then $\omega_{eff}\to-1$. Figure 11b depicts the variation of effective EoS parameter over statefinder $s(z)$, that reveals that the EoS $\omega_{eff}$ is directly proportional to $r(z)$. From Eqs.~\eqref{eq26} \& \eqref{eq28}, we can obtain the relationship between $\omega_{eff}$ \& $s$ as $3As\omega_{eff}+2B\omega_{eff}+2B=0$ for model-I. Also, we can express it as $\omega_{eff}=-\frac{2B}{3As+2B}$ that represents as $s\to0$ then $\omega_{eff}\to-1$ and as $s\to\infty$ then $\omega_{eff}\to0$.
	%%%%%%%%%%%%%%%%%%%%%%%%%%%%%%%%%%%%%%%%%%%%%%%%%%%%%%%%%%%%
	%%%%%%%%%%%%%%%%%%%%%%%%%%%%%%%%%%%% Figure 12
	%%%%%%%%%%%%%%%%%%%%%%%%%%%%%%%%%%%%%%%%%%%%%%%%%%%%%%%%%%%%
	\begin{figure}[H]
		\centering
		a.\includegraphics[width=8cm,height=6cm,angle=0]{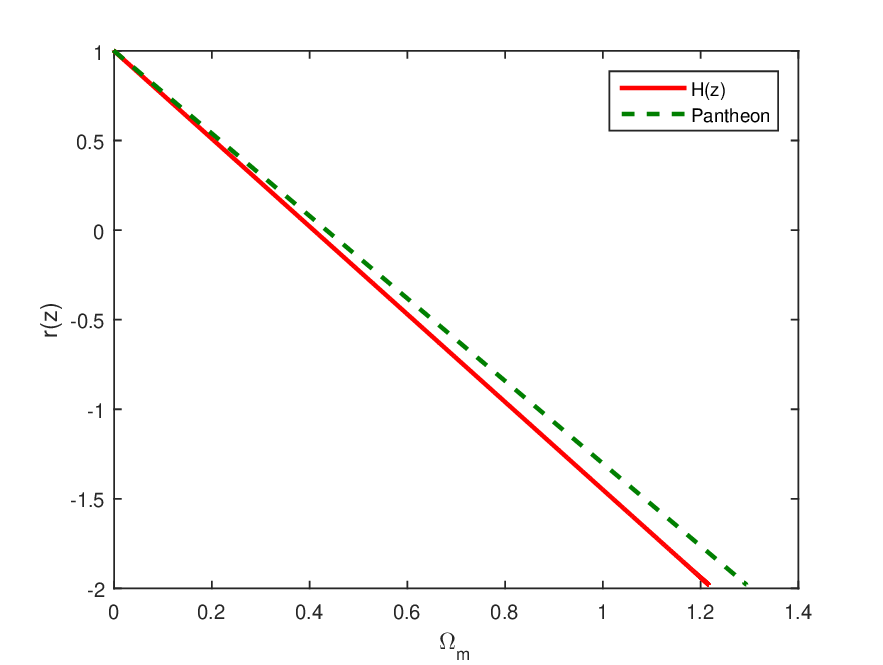}
		b.\includegraphics[width=8cm,height=6cm,angle=0]{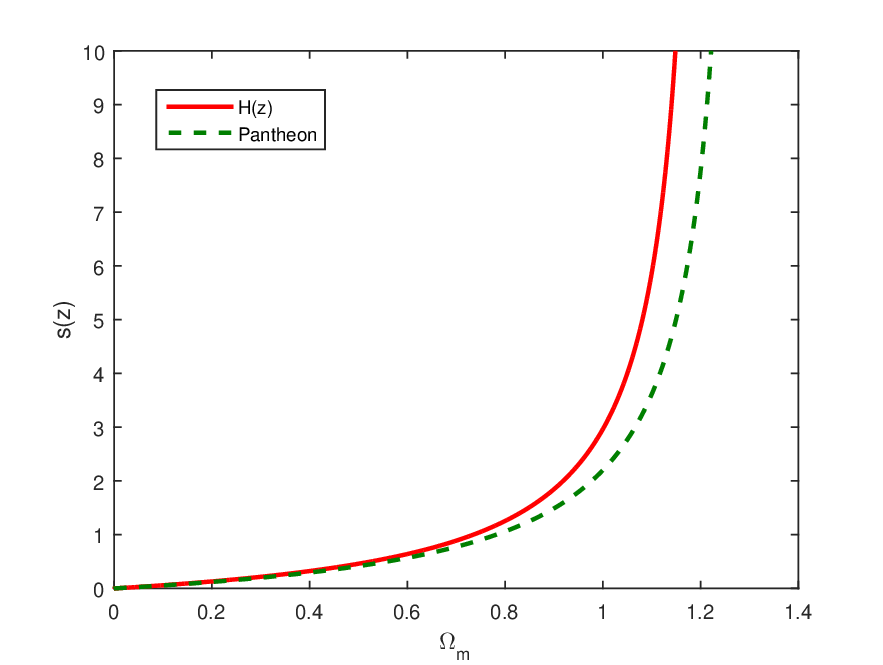}
		\caption{The plots of statefinder parameters $r(z)$ \& $s(z)$ versus matter energy density parameter $\Omega_{m}$, respectively.}
	\end{figure}
	%%%%%%%%%%%%%%%%%%%%%%%%%%%%%%%%%%%%%%%%%%%%%%%%%%%%%%%%%%%%%%%%%%
	Figure 12a depicts the variation of statefinder parameter $r(z)$ over $\Omega_{m}$ that reveals the relationship between them. From Eqs.~\eqref{eq25} \& \eqref{eq30}, we can derive the relationship between $r$ \& $\Omega_{m}$ as $r=1-C\Omega_{m}$, $C=\frac{3k^{2}}{\rho_{m0}\kappa^{2}}$. Figure 12b shows the variation of statefinder $s$ over matter energy density parameter $\Omega_{m}$ that reveals the relationship between them. The relationship between $s$ \& $\Omega_{m}$ is derived from Eqs.~\eqref{eq26} \& \eqref{eq30} as $3k^{2}s\Omega_{m}+2k^{2}\Omega_{m}-3\rho_{m0}\kappa^{2}s=0$ for model-I.
	%%%%%%%%%%%%%%%%%%%%%%%%%%%%%%%%%%%%%%%%%%%%%%%%%%%%%%%%%%%%%%%%
\subsection{Model-II}
%%%%%%%%%%%%%%%%%%%%%%%%%%%%%%%%%%%%%%%%%%%%%%%%%%%%%%%%%%%%%%%%
	In the second model, we have obtained the scale factor by solving the field equation \eqref{eq32} and obtained the scale factor $a(t)$ as in Eq.~\eqref{eq35}. The geometrical evolution of $a(t)$ for model-II is shown in figure 13a with cosmic time $t$ for two observational datasets $H(z)$ and Pantheon SNe Ia. At present, we have assumed the value of scale factor as $a_{0}=a_{t_{0}}=1$, where $t_{0}$ denotes the present age of the universe. We have estimated the present age of the universe as $t_{0}=0.01381_{-0.00008}^{+0.00005}, 0.01365_{-0.00006}^{+0.00002}$ or $t_{0}=13.50_{-0.07824}^{+0.04890}, 13.29_{-0.05868}^{+0.01956}$ Gyrs, respectively for two observational datasets. At $t=0$, $a(0)=0$ and as $t\to\infty$, then $a(t)\to\infty$, this shows the expansion of the universe.
	%%%%%%%%%%%%%%%%%%%%%%%%%%%%%%%%%%%%%%%%%%%%%%%%%%%%%%%%%%%%
	%%%%%%%%%%%%%%%%%%%%%%%%%%%%%%%%%%%% Figure 13
	%%%%%%%%%%%%%%%%%%%%%%%%%%%%%%%%%%%%%%%%%%%%%%%%%%%%%%%%%%%%
	\begin{figure}[H]
		\centering
		a.\includegraphics[width=8cm,height=6cm,angle=0]{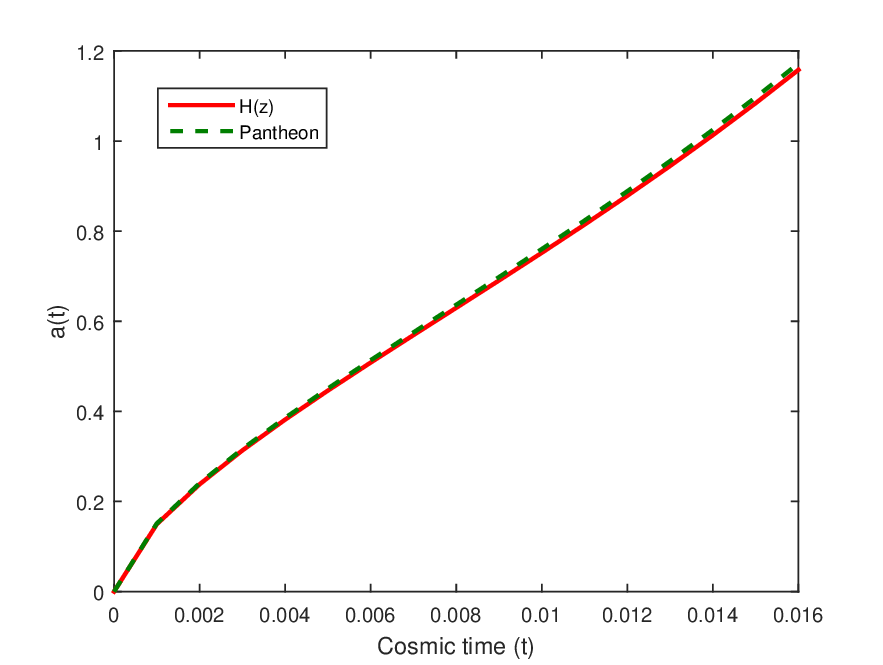}
		b.\includegraphics[width=8cm,height=6cm,angle=0]{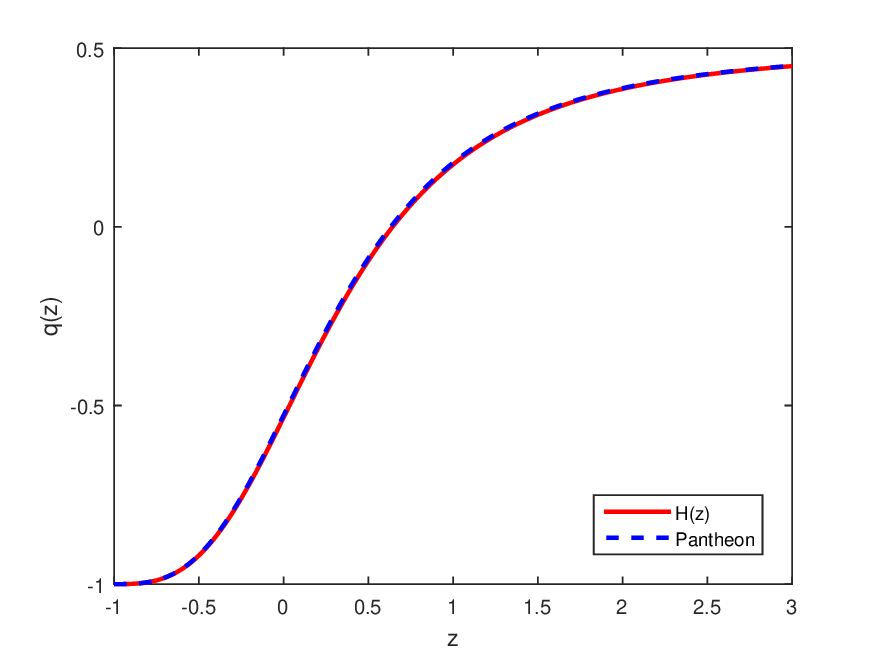}
		\caption{The plot of scale factor $a(t)$ versus cosmic time $t$, and the deceleration parameter $q(z)$ versus redshift $z$, respectively.}
	\end{figure}
	%%%%%%%%%%%%%%%%%%%%%%%%%%%%%%%%%%%%%%%%%%%%%%%%%%%%%%%%%%%%%%%%%%
	The equation \eqref{eq37} represents the mathematical expression for the deceleration parameter derived from the scale factor \eqref{eq35}. We have obtained the expression of $q$ in terms of redshift $z$ as in Eq.~\eqref{eq39}. The geometrical behaviour of $q(z)$ is represents in figure 12b, and from the figure 13b, one can see that $q(z)$ is an increasing function of $z$. The evolution of $q(z)$ shows a signature-flipping point (transition) denoted by $z_{t}$. We have estimated the present values of transition redshift as $z_{t}=0.642_{-0.001}^{+0.003}, 0.633_{-0.001}^{+0.012}$, respectively for two observational datasets $H(z)$ and Pantheon SNe Ia, which are in good agreement with recent observed values of $z_{t}$ in \cite{ref82}-\cite{ref87}. At $z=0$, the value of deceleration parameter is called present value and it is denoted by $q_{0}$. We have estimated the present value of deceleration parameter $q$ as $q_{0}=-0.5338\pm0.0047, -0.5275\pm0.0063$, respectively along two considered datasets. Thus the value $q_{0}<0$ reveals that our universe model is accelerating at present. From figure 13b, we can see that $q\to-1$ as $z\to-1$ and $q\to0.5$ for $z\to\infty$ that depicts the existence of such universe which is early decelerating and late-time accelerating expanding in nature. The transition redshift is generally obtained as
		\begin{equation}\label{eq51}
		z_{t}=\left[\frac{3(k_{2}+\lambda k_{4})}{k^{2}(3+k_{1}+3\lambda)}\right]^{\frac{1}{3}}-1.
	\end{equation}
		For the model-II, the mathematical expression for matter energy density parameter $\Omega_{m}$ and dark energy density parameter $\Omega_{F}$ are shown by the Eqs.~\eqref{eq43} \& \eqref{eq44}, respectively. The geometrical evolution of $\Omega_{m}$ \& $\Omega_{F}$ over $z$ are shown in figure 14a \& 14b, respectively. From figure 14a \& 14b, one can see that as $z\to-1$, then $\Omega_{m}\to0$ \& $\Omega_{F}\to1$ that represents the $\Lambda$CDM tendency of the model, on the other hand, as $z\to\infty$, then $\Omega_{m}$ tends to $1.4$ and $\Omega_{F}\to-\lambda-0.5k_{1}$ that reveals the early matter dominated universe. We have estimated the present values of these parameters as $\Omega_{m0}=0.4167_{-0.0073}^{+0.0024}, 0.4160_{-0.0130}^{+0.0130}$ and $\Omega_{F0}=0.5719_{-0.1198}^{+0.1080}, 0.5655_{-0.0456}^{+0.0546}$, respectively for two observational datasets $H(z)$ and Pantheon SNe Ia.
	%%%%%%%%%%%%%%%%%%%%%%%%%%%%%%%%%%%%%%%%%%%%%%%%%%%%%%%%%%%%
	%%%%%%%%%%%%%%%%%%%%%%%%%%%%%%%%%%%% Figure 14
	%%%%%%%%%%%%%%%%%%%%%%%%%%%%%%%%%%%%%%%%%%%%%%%%%%%%%%%%%%%%
	\begin{figure}[H]
		\centering
		a.\includegraphics[width=8cm,height=6cm,angle=0]{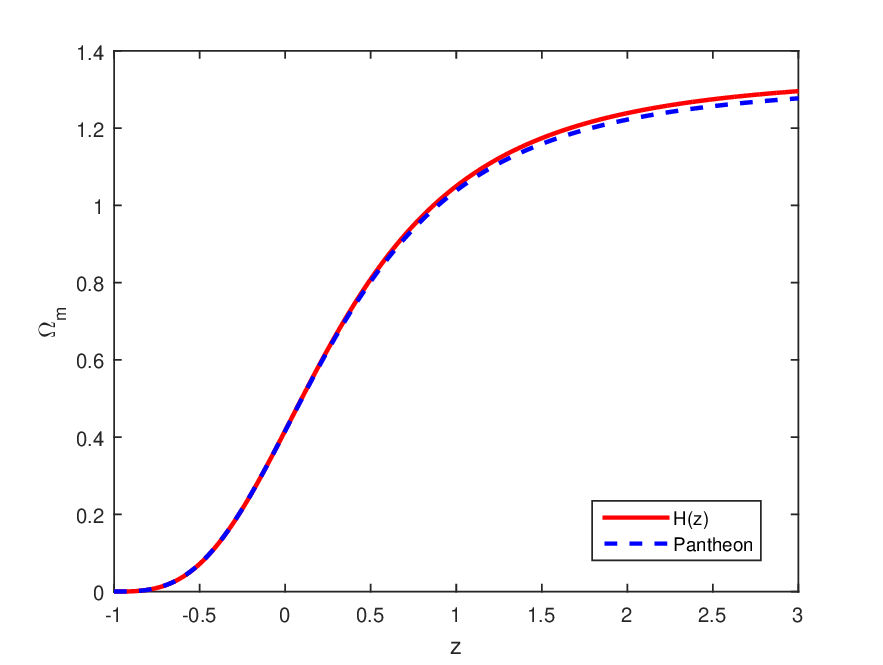}
		b.\includegraphics[width=8cm,height=6cm,angle=0]{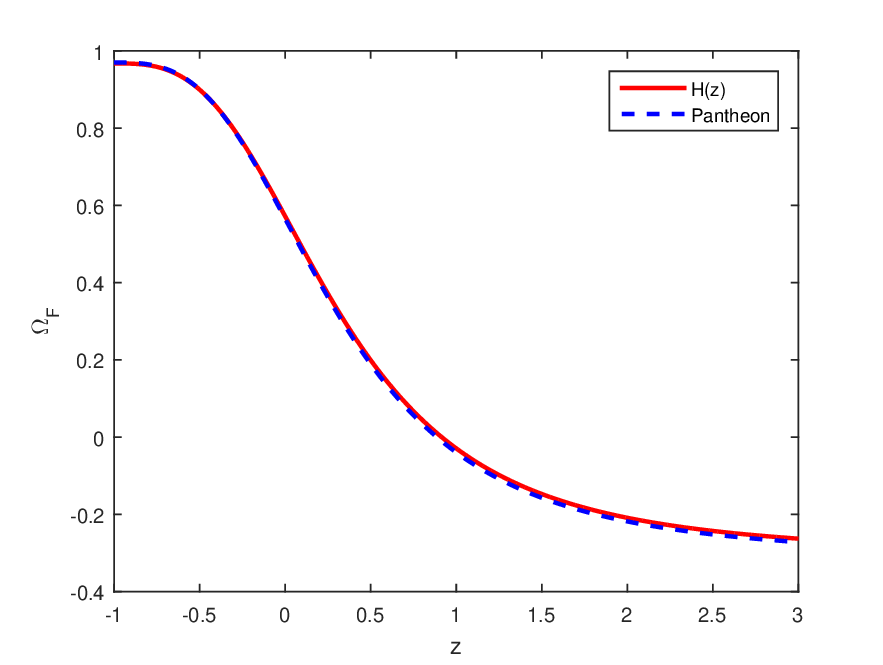}
		\caption{The plots of total energy density parameters $\Omega_{m}$ and $\Omega_{F}$ versus redshift $z$, respectively.}
	\end{figure}
	%%%%%%%%%%%%%%%%%%%%%%%%%%%%%%%%%%%%%%%%%%%%%%%%%%%%%%%%%%%%%%%%%%
    %%%%%%%%%%%%%%%%%%%%%%%%%%%%%%%%%%%%%%%%%%%%%%%%%%%%%%%%%%%%
	%%%%%%%%%%%%%%%%%%%%%%%%%%%%%%%%%%%% Figure 15
	%%%%%%%%%%%%%%%%%%%%%%%%%%%%%%%%%%%%%%%%%%%%%%%%%%%%%%%%%%%%
	\begin{figure}[H]
		\centering
		a.\includegraphics[width=8cm,height=6cm,angle=0]{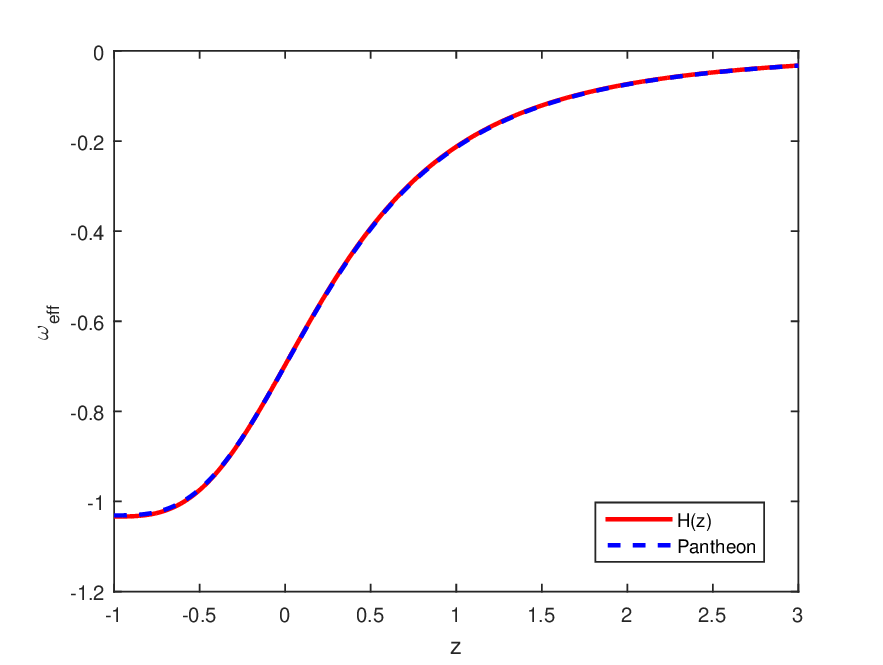}
		b.\includegraphics[width=8cm,height=6cm,angle=0]{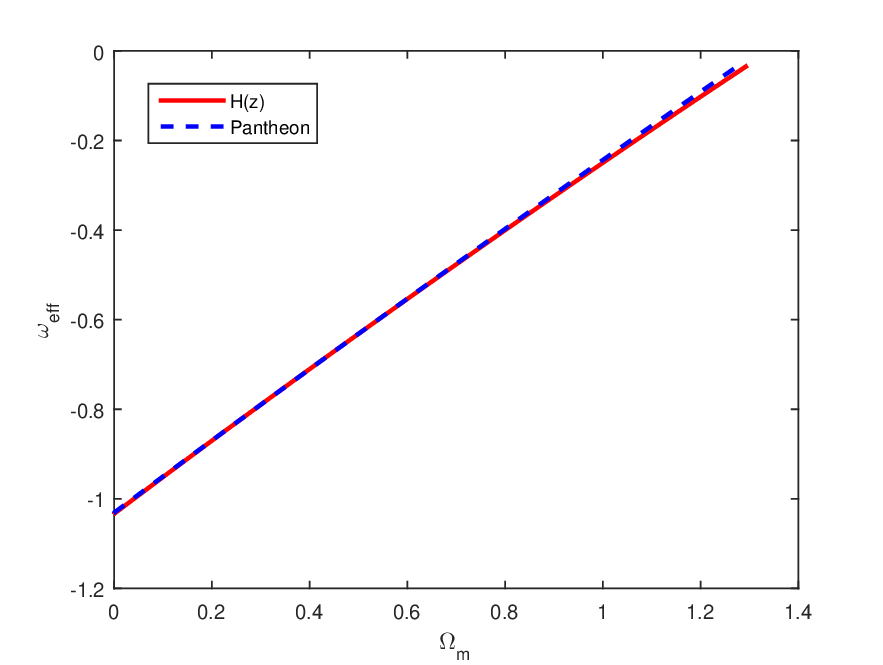}
		\caption{The variation of effective EoS parameter $\omega_{eff}$ versus $z$ \& $\Omega_{m}$, respectively.}
	\end{figure}
	%%%%%%%%%%%%%%%%%%%%%%%%%%%%%%%%%%%%%%%%%%%%%%%%%%%%%%%%%%%%%%%%%%
	For model-II, the mathematical expression of effective EoS parameter $\omega_{eff}$ is obtained as in Eq.~\eqref{eq42}, and its geometrical evolution over $z$ is shown in figure 15a. From figure 15a, one can see that $\omega_{eff}\to-1.031$ as $z\to-1$, and it crosses the $\Lambda$CDM value $\omega_{eff}=-1$ at $z=-0.599, -0.6$, respectively for two datasets. On the other hand, $\omega_{eff}\to0$ as $z\to\infty$ that represents that early universe was matter dominated. The present value of effective EoS parameter is measured as $\omega_{eff}=-0.6971_{-0.1066}^{+0.0696}, -0.6979_{-0.0440}^{+0.0287}$, respectively along two observational datasets $H(z)$ and Pantheon SNe Ia. Figure 15b represents the variation of $\omega_{eff}$ versus $\Omega_{m}$, and one can see that they have relationship which is obtained from Eqs.~\eqref{eq42} \& \eqref{eq44} as $(A+B)k^{2}\Omega_{m}\omega_{eff}-B\rho_{m0}\kappa^{2}\omega_{eff}+(B-n_{2}k_{1}k^{2})k^{2}\Omega_{m}-B\rho_{m0}\kappa^{2}+n_{2}\rho_{m0}k_{1}k^{2}\kappa^{2}=0$ where $A=3n_{2}(2\kappa^{2}\rho_{m0}-k_{1}k^{2}-2\lambda k^{2})$ and $B=3k^{2}(2n_{2}\lambda+n_{2}k_{1}-k_{2}-k_{4}\lambda)$. It can be also expressed as
	\begin{equation}\label{eq52}
		\omega_{eff}=\frac{B\rho_{m0}\kappa^{2}-n_{2}\rho_{m0}k_{1}k^{2}\kappa^{2}-(B-n_{2}k_{1}k^{2})k^{2}\Omega_{m}}{(A+B)k^{2}\Omega_{m}-B\rho_{m0}\kappa^{2}}
	\end{equation}
	From Eq.~\eqref{eq52}, we can obtained for $\Omega_{m}\to0$, $\omega_{eff}\to-1-\frac{k_{1}}{6-k_{1}}, k_{1}<6$ that gives cosmological constant value $\omega_{eff}=-1$ for $k_{1}=0$, phantom and super-phantom value for $k_{1}<6$.\\
		The mathematical expressions of statefinder parameters for model-II are represented by the Eqs.~\eqref{eq40} \& \eqref{eq41}, respectively, and their geometrical evolution over $z$ are shown in figure 16a \& 16b. From figure 16a \& 16b, one can observe that $(s, r)\to(0,1)$ as $z\to-1$ that reveals that model obtained $\Lambda$CDM stage at late-time universe. We have measured the present values of these parameters as $r_{0}=0.0676_{-0.0094}^{+0.0094}, 0.0550_{-0.0126}^{+0.0126}$ and $s_{0}=0.3007_{-0.0055}^{+0.0055}, 0.3066_{-0.006}^{+0.006}$, respectively for two datasets.
	%%%%%%%%%%%%%%%%%%%%%%%%%%%%%%%%%%%%%%%%%%%%%%%%%%%%%%%%%%%%
	%%%%%%%%%%%%%%%%%%%%%%%%%%%%%%%%%%%% Figure 16
	%%%%%%%%%%%%%%%%%%%%%%%%%%%%%%%%%%%%%%%%%%%%%%%%%%%%%%%%%%%%
	\begin{figure}[H]
		\centering
		a.\includegraphics[width=8cm,height=6cm,angle=0]{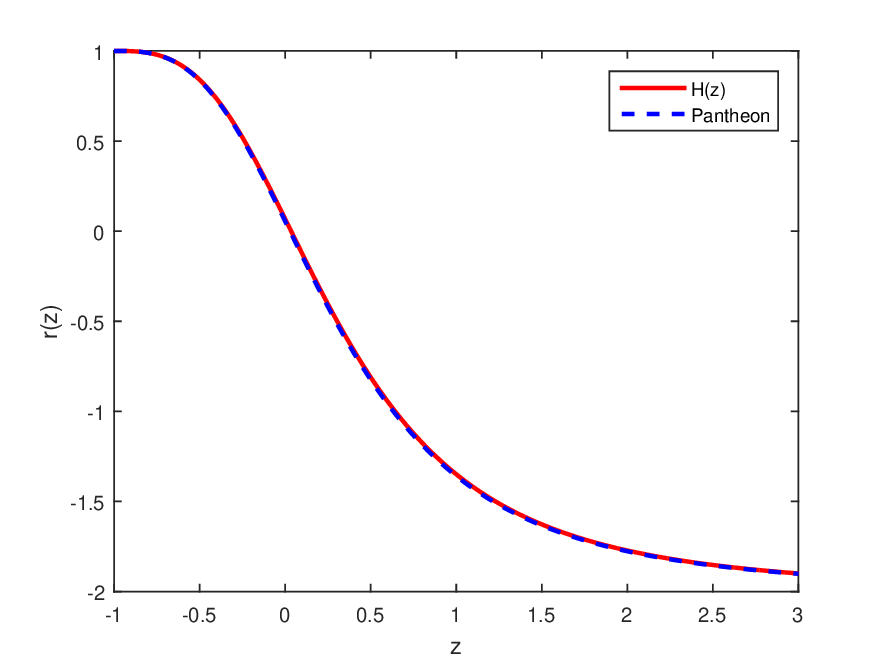}
		b.\includegraphics[width=8cm,height=6cm,angle=0]{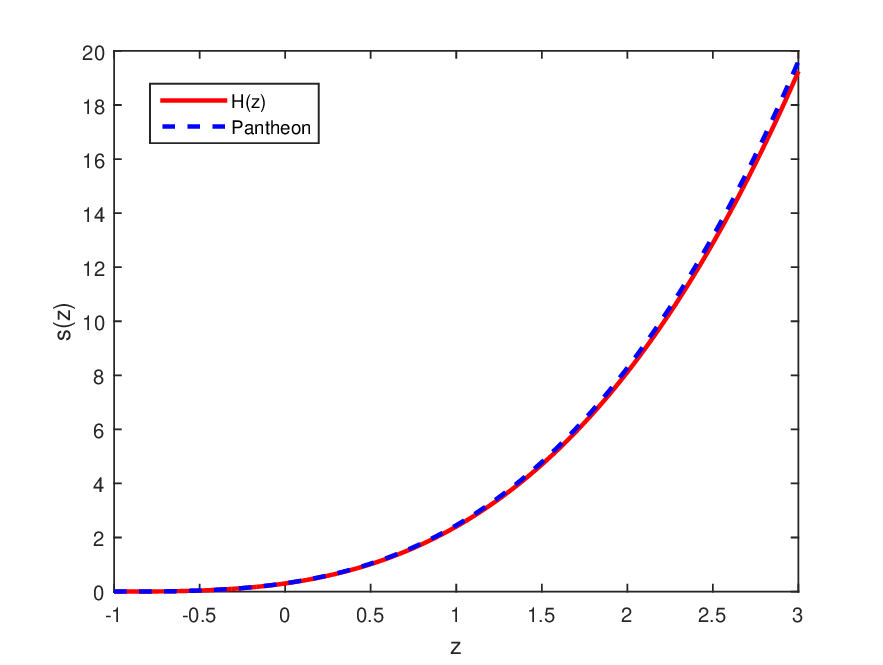}
		\caption{The plots of statefinder diagnostic parameters $r(z)$ \& $s(z)$ versus $z$, respectively.}
	\end{figure}
	%%%%%%%%%%%%%%%%%%%%%%%%%%%%%%%%%%%%%%%%%%%%%%%%%%%%%%%%%%%%%%%%%%
	%%%%%%%%%%%%%%%%%%%%%%%%%%%%%%%%%%%%%%%%%%%%%%%%%%%%%%%%%%%%
	%%%%%%%%%%%%%%%%%%%%%%%%%%%%%%%%%%%% Figure 17
	%%%%%%%%%%%%%%%%%%%%%%%%%%%%%%%%%%%%%%%%%%%%%%%%%%%%%%%%%%%%
	\begin{figure}[H]
		\centering
		a.\includegraphics[width=8cm,height=6cm,angle=0]{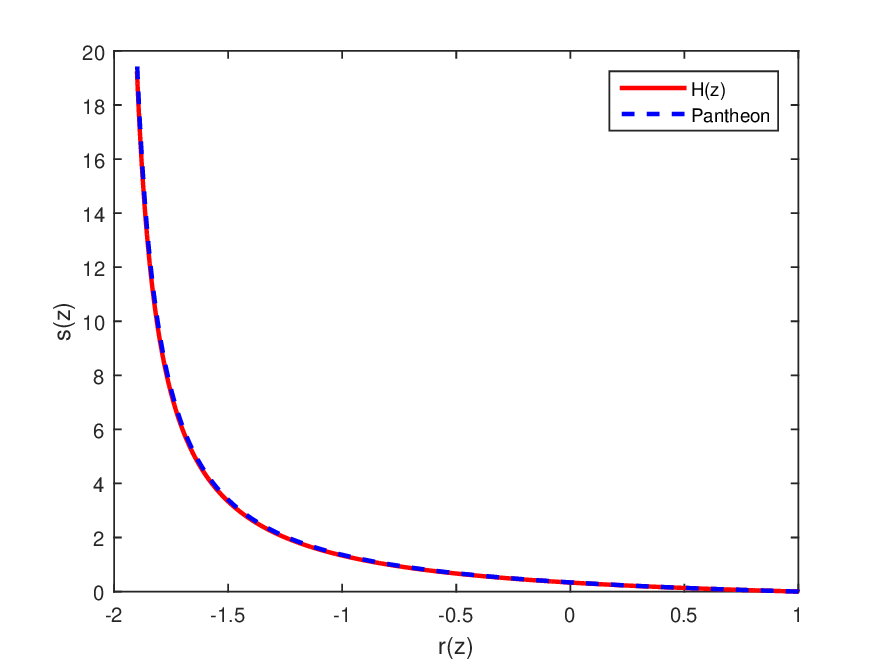}
		b.\includegraphics[width=8cm,height=6cm,angle=0]{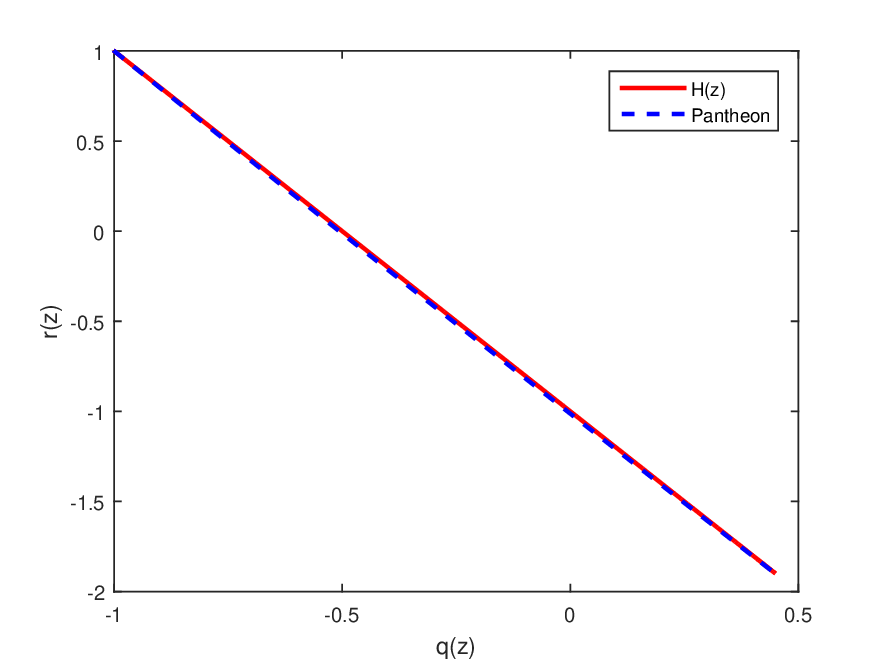}
		\caption{The plots of statefinder $s(z)$ versus $r(z)$, and statefinder $r(z)$ versus deceleration parameter $q(z)$, respectively.}
	\end{figure}
	%%%%%%%%%%%%%%%%%%%%%%%%%%%%%%%%%%%%%%%%%%%%%%%%%%%%%%%%%%%%%%%%%%
	%%%%%%%%%%%%%%%%%%%%%%%%%%%%%%%%%%%%%%%%%%%%%%%%%%%%%%%%%%%%
	%%%%%%%%%%%%%%%%%%%%%%%%%%%%%%%%%%%% Figure 18
	%%%%%%%%%%%%%%%%%%%%%%%%%%%%%%%%%%%%%%%%%%%%%%%%%%%%%%%%%%%%
	\begin{figure}[H]
		\centering
		a.\includegraphics[width=8cm,height=6cm,angle=0]{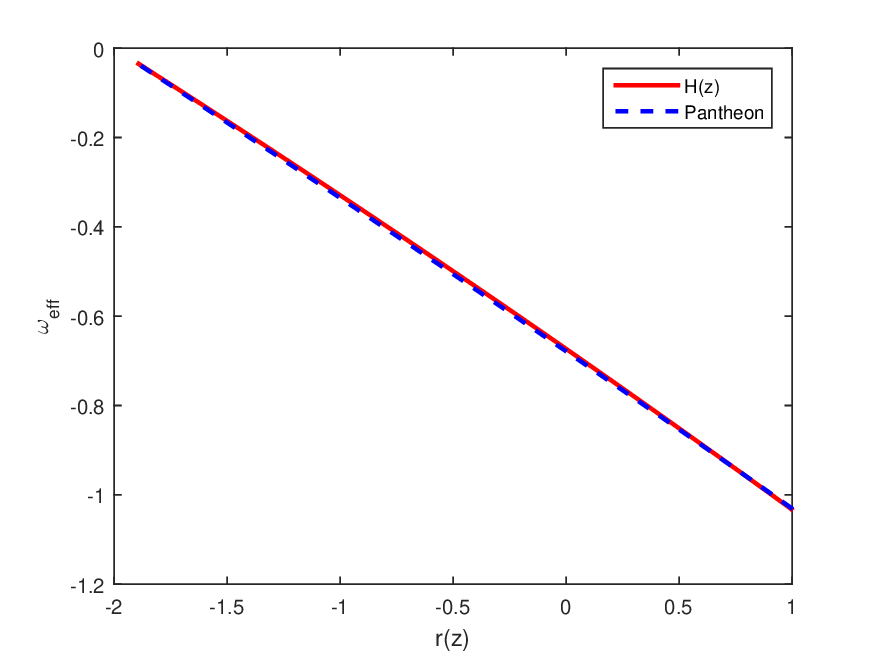}
		b.\includegraphics[width=8cm,height=6cm,angle=0]{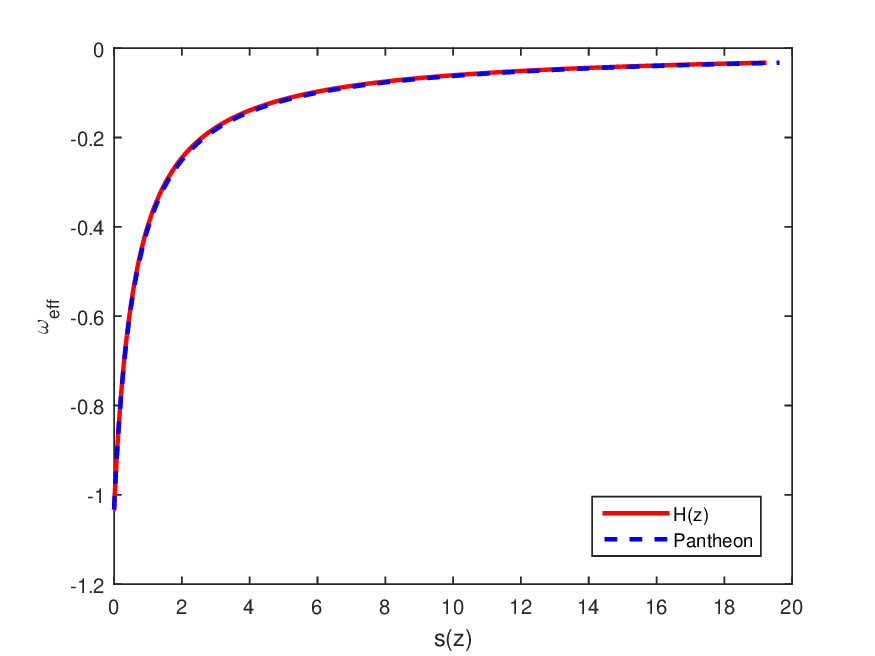}
		\caption{The plots of effective EoS parameter versus statefinder parameters $r(z)$ \& $s(z)$, respectively.}
	\end{figure}
	%%%%%%%%%%%%%%%%%%%%%%%%%%%%%%%%%%%%%%%%%%%%%%%%%%%%%%%%%%%%%%%%%%
		Figure 17a represents the relationship between statefinder parameters $r, s$ for the model-II, and figure 17b represents the relationship between statefinder $r(z)$ and deceleration parameter $q(z)$. From figure 17a \& 17b, one can observe that as $r\to1$, then $s\to0$, while $r\to1$ when $q\to-1$.\\
		
	Figure 18a \& 18b represent the relationship between effective EoS parameters $\omega_{eff}$ and statefinder parameters $r, s$. From Eqs.~\eqref{eq40}, \eqref{eq41} \& \eqref{eq42}, we obtain these relationship as $(A+C)r\omega_{eff}+(2C-A)\omega_{eff}+(C-B)r+2C-2B=0$, and $3As\omega_{eff}-2C\omega_{eff}-2C+2B=0$ where $A=3n_{2}(2\kappa^{2}\rho_{m0}-k_{1}k^{2}-2\lambda k^{2})$, $B=n_{2}k_{1}k^{2}$ and $C=3k^{2}(2n_{2}\lambda+n_{2}k_{1}-k_{2}-k_{4}\lambda)$. These relations can be expressed as
	\begin{equation}\label{eq53}
		\omega_{eff}=\frac{(B-C)(r+2)}{(A+B)r+(2C-A)},~~~~\omega_{eff}=\frac{2(C-B)}{3As-2C}
	\end{equation}
	For $r\to1$ \& $s\to0$, we find respectively the effective EoS parameter as $\omega_{eff}\to\frac{3(B-C)}{B+2C}$ and $\omega_{eff}\to-1+\frac{B}{C}$.
	%%%%%%%%%%%%%%%%%%%%%%%%%%%%%%%%%%%%%%%%%%%%%%%%%%%%%%%%%%%%
	%%%%%%%%%%%%%%%%%%%%%%%%%%%%%%%%%%%% Figure 19
	%%%%%%%%%%%%%%%%%%%%%%%%%%%%%%%%%%%%%%%%%%%%%%%%%%%%%%%%%%%%
	\begin{figure}[H]
		\centering
		a.\includegraphics[width=8cm,height=6cm,angle=0]{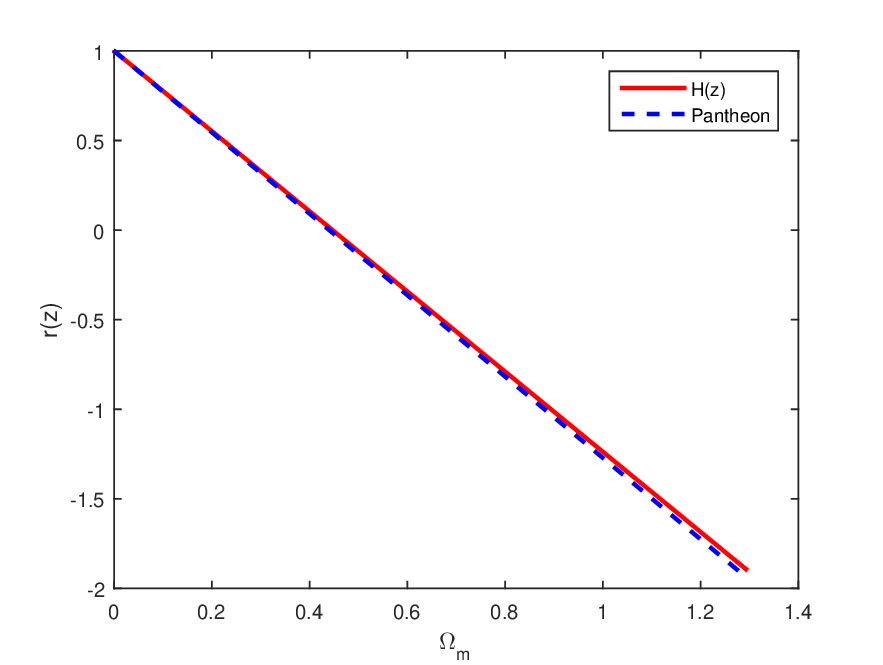}
		b.\includegraphics[width=8cm,height=6cm,angle=0]{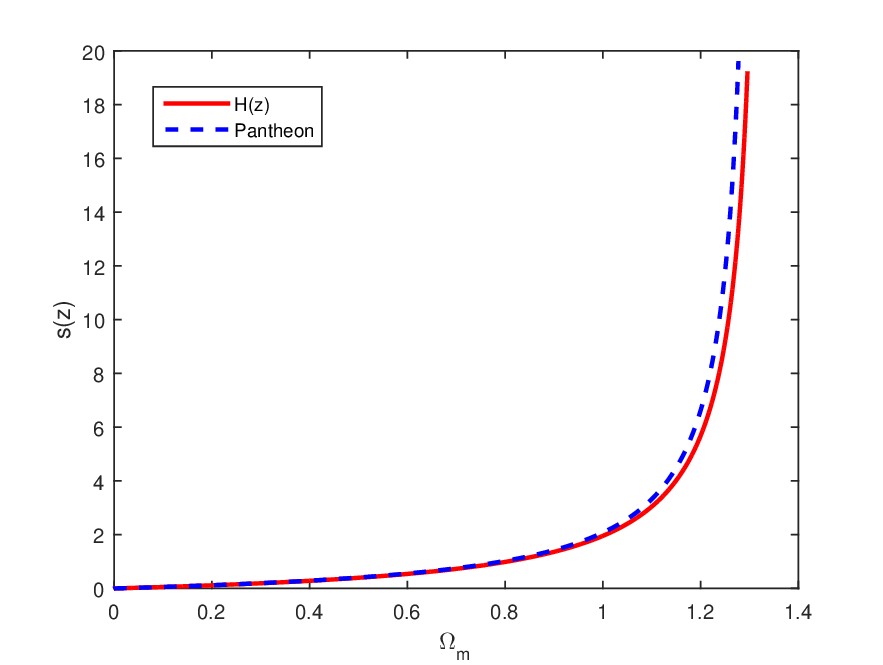}
		\caption{The plots of statefinder parameters $r(z)$ \& $s(z)$ versus $\Omega_{m}$, respectively.}
	\end{figure}
	%%%%%%%%%%%%%%%%%%%%%%%%%%%%%%%%%%%%%%%%%%%%%%%%%%%%%%%%%%%%%%%%%%
	Figure 19a \& 19b represent the relationship between statefinder parameters $r, s$ and matter energy density parameter $\Omega_{m}$. We can derive these relationship from the Eqs.~\eqref{eq40}, \eqref{eq41} \& \eqref{eq44} as $r=1-\frac{3k^{2}}{\rho_{m0}\kappa^{2}}\Omega_{m}$ and $3k^{2}s\Omega_{m}-3\rho_{m0}\kappa^{2}s+2k^{2}\Omega_{m}=0$.\\
	
	Thus, both the model-I \& model-II are able to explain the late-time accelerating as well as early decelerating scenario of the universe but it does not able to explain the early time accelerating (inflation) scenario of the universe. In the model-II, the effective EoS parameter crosses the cosmological constant value $\omega_{eff}=-1$ while model-I does not so and it tends to cosmological constant value $-1$.
	%%%%%%%%%%%%%%%%%%%%%%%%%%%%%%%%%%%%%%%%%%%%%%%%%%%%%%%%%%%%%%%%
	\section{Conclusions}
	%%%%%%%%%%%%%%%%%%%%%%%%%%%%%%%%%%%%%%%%%%%%%%%%%%%%%%%%%%%%%%%%
	
	     Using a flat Friedmann-Lematre-Robertson-Walker (FLRW) spacetime metric, we have studied various exact cosmological models in Myrzakulov gravity in this study. We have taken into consideration the modified Lagrangian function $F(R,T)=R+\lambda T$, where $\lambda$ is a model parameter and $R, T$ are the Ricci curvature scalar and the torsion scalar with regard to non-special connection, respectively. We have found two exact solutions for the factor of scale $a(t)$ in two distinct scenarios. We have derived a number of geometrical parameters to study the universe's cosmological properties by using this scale factor. With $1-\sigma, 2-\sigma$, and $3-\sigma$ areas, we have determined the best fit values of model parameters using MCMC analysis on two types of recent observational datasets, such as $H(z)$ and Pantheon SNe Ia samples. We have studied geometrical and cosmological parameters in a relativistic comparison manner. The effective equation of state (EoS) parameter, $\omega_{eff}$, varies as $-1.031\le\omega_{eff}\le0$ in model-II, whereas it varies in the range $-1\le\omega_{eff}\le0$ in model-I. The main features of the derived models are as follows:
	     \begin{itemize}
	     	\item Both model are transit phase expanding universe with transition redshift in the range $0.6<z_{t}<0.8$ which is in good agreement with recent measurement \cite{ref82}-\cite{ref87}.
	     	\item Both models are decelerating to accelerating phase of expansion and the present values of $q(z)$ varies over $-0.6<q_{0}<-0.5$ which are in good agreement with recent observations \cite{ref1}-\cite{ref7}.
	     	\item Both models are able to explain the late-time accelerating as well as early decelerating scenario of the universe but it does not able to explain the early time accelerating (inflation) scenario of the universe.
	     	\item The effective EoS parameter in Model-I evolves as $-1\le\omega_{eff}\le0$ over $-1\le z \le \infty$.
	     	\item In Model-II, the effective EoS parameter varies as $-1.031\le\omega_{eff}\le0$ over $-1\le z \le \infty$.
	     	\item We have found that the effective EoS parameter $\omega_{eff}$ is related to statefinder parameters $r, s$ and their behaviour is in acceptable range.
	     	\item Model-II evolves from matter dominated stage, passes through quintessence, cosmological constant and tends to phantom and super-phantom stages of the expanding universe.
	     	\item While Model-I evolves from matter dominated, passes to quintessence and tends to $\Lambda$CDM model.
	     	\item We have found good relationship among the parameters $\omega_{eff}$, $\Omega_{m}$, $q, r$ \& $s$.
	     	\item We have found the value of Hubble constant $H_{0}=67.967_{-0.078202}^{+0.063326}, 68.193_{-0.12074}^{+0.10342}$ Km/s/Mpc, respectively for two models.
	     	\item We have found the present age of the universe as $t_{0}=13.33_{-0.19}^{+0.26}, 13.48_{-0.05}^{+0.02}$ Gyrs, respectively along two datasets $H(z)$ and Pantheon SNe Ia for model-I, while for model-II, it is found as $t_{0}=13.50_{-0.07824}^{+0.04890}, 13.29_{-0.05868}^{+0.01956}$ Gyrs, respectively for two observational datasets
	     \end{itemize}
	     
	     Thus, we can conclude that the choice of non-special connection (i.e. in resultant the choice of $u$ \& $v$) has an important role in the dynamical dark energy evolution history of the universe. Therefore, it needs further investigation and attracts cosmologists to review it.

	%%%%%%%%%%%%%%%%%%%%%%%%%%%%%%%%%%%%%%%%%%%%%%%%%%%%%%%%%%%%%%%%
	\section{Acknowledgments}
	%%%%%%%%%%%%%%%%%%%%%%%%%%%%%%%%%%%%%%%%%%%%%%%%%%%%%%%%%%%%%%%%
	This work was supported by the Ministry of Science and Higher Education of the Republic of Kaza-
khstan, GrantAP14870191.

	    %%%%%%%%%%%%%%%%%%%%%%%%%%%%%%%%%%%%%%%%%%%%%%%%%%%
	

\begin{thebibliography}{}
		%%%%%%%%%%%%%%%%%%%%%%%%%%%%%%%%%%%%%%%%%%%%%%%%%%%%
		\bibitem {ref1}
		A. G. Riess, {\it et al.}, Observational evidence from supernovae for an accelerating universe and a cosmological constant, \textit{Astron. J.} \textbf{116}, 1009 (1998).
		\bibitem {ref2}
		S. Perlmutter, {\it et al.}, Measurements of omega and lambda from $42$ high-redshift supernovae, \textit{Astrophys. J.} \textbf{517}, 565 (1999).
		\bibitem {ref3}
		A. G. Riess, \textit{et al.}, Type-Ia supernova discoveries of $z\geq1$ from the Hubble space telescope: Evidence from past deceleration and constraints on dark energy evolution, \textit{Astrophys. J.} \textbf{607}, 665 (2004).
		\bibitem {ref4}
		D. J. Eisenstein, \textit{et al.}, Detection of the baryon acoustic peak in the large-scale correlation function of SDSS luminous red galaxies, \textit{Astrophys. J.} \textbf{633}, 560 (2005).
		\bibitem {ref5}
		W. J. Percival, \textit{et al.}, Baryon acoustic oscillations in the Sloan Digital Sky Survey data release 7 galaxy sample, \textit{Mon. Not. R. Astron. Soc.} \textbf{401}, 2148 (2010).
		\bibitem {ref6}
		D. N. Spergel, \textit{et al.}, First-year Wilkinson Microwave Anisotropy Probe (WMAP)
		observations: Determination of Cosmological parameters, \textit{Astrophys. J. Suppl. Ser.} \textbf{148}, 175 (2003). astro-ph/0302209.
		\bibitem {ref7}
		T. Koivisto and D. F. Mota, Dark energy anisotropic stress and large scale structure formation, \textit{Phys. Rev. D} \textbf{73}, 083502 (2006).
		\bibitem {ref8}
		E. J. Copeland, M. Sami and S. Tsujikawa, Dynamics of dark energy, \textit{Int. J. Mod. Phys. D} \textbf{15}, 1753
		(2006) [arXiv:hep-th/0603057].
		\bibitem {ref9}
		Y. -F. Cai, E. N. Saridakis, M. R. Setare and J. -Q. Xia, Quintom Cosmology: Theoretical implications and observations, \textit{Phys. Rept.} \textbf{493}, 1 (2010) [arXiv:0909.2776].
		\bibitem {ref10}
		K. A. Olive, Inflation, \textit{Phys. Rept.} \textbf{190}, 307 (1990).
		\bibitem {ref11}
		N. Bartolo, E. Komatsu, S. Matarrese and A. Riotto, Non-Gaussianity from inflation: Theory and observations, \textit{Phys. Rept.} \textbf{402}, 103 (2004) [arXiv:astro-ph/0406398].
		\bibitem {ref12}
		S. Capozziello and M. De Laurentis, Extended Theories of Gravity, \textit{Phys. Rept.} \textbf{509}, 167 (2011)
		[arXiv:1108.6266].
		\bibitem {ref13}
		Y. F. Cai, S. Capozziello, M. De Laurentis and E. N. Saridakis, $f(T)$ teleparallel gravity and cosmology, \textit{Rept. Prog. Phys.} \textbf{79}, 106901 (2016) [arXiv:1511.07586].
		\bibitem {ref14}
		P. Brax, C. van de Bruck and A. C. Davis, Brane world cosmology, \textit{Rept. Prog. Phys.} \textbf{67}, 2183-2232
		(2004) [arXiv:hep-th/0404011].
		\bibitem {ref15}
		A. De Felice and S. Tsujikawa, $f(R)$ theories, \textit{Living Rev. Rel.} \textbf{13}, 3 (2010) [arXiv:1002.4928].
		\bibitem {ref16}
		S. Nojiri and S. D. Odintsov, Unified cosmic history in modified gravity: from $F(R)$ theory to	Lorentz non-invariant models, \textit{Phys. Rept.} \textbf{505}, 59 (2011) [arXiv:1011.0544].
		\bibitem {ref17}
		S. Nojiri and S. D. Odintsov, Modified Gauss-Bonnet theory as gravitational alternative for dark energy, \textit{Phys. Lett. B} \textbf{631}, 1 (2005) [arXiv:hep-th/0508049].
		\bibitem {ref18}
		A. De Felice and S. Tsujikawa, Construction of cosmologically viable $f(G)$ dark energy models,	\textit{Phys. Lett. B} \textbf{675}, 1 (2009) [arXiv:0810.5712].
		\bibitem {ref19}
		D. Lovelock, The Einstein tensor and its generalizations, \textit{J. Math. Phys.} \textbf{12}, 498 (1971).
		\bibitem {ref20}
		N. Deruelle and L. Farina-Busto, The Lovelock Gravitational Field Equations in Cosmology, \textit{Phys. Rev. D} \textbf{41}, 3696 (1990).
		\bibitem {ref21}
		G. W. Horndeski, Second-order scalar-tensor field equations in a four-dimensional space, \textit{Int. J. Theor. Phys.} \textbf{10}, 363-384 (1974).
		\bibitem {ref22}
		A. Nicolis, R. Rattazzi and E. Trincherini, The Galileon as a local modification of gravity, \textit{Phys. Rev. D} \textbf{79}, 064036 (2009) [arXiv:0811.2197].
		\bibitem {ref23}
		C. Deffayet, G. Esposito-Farese and A. Vikman, Covariant Galileon, \textit{Phys. Rev. D} \textbf{79}, 084003
		(2009) [arXiv:0901.1314].
		\bibitem {ref24}
		R. Ferraro and F. Fiorini, Modified teleparallel gravity: Inflation without inflaton, \textit{Phys. Rev. D} \textbf{75},
		084031 (2007) [arXiv:gr-qc/0610067].
		\bibitem {ref25}
		E. V. Linder, Einstein's Other Gravity and the Acceleration of the Universe, \textit{Phys. Rev. D} \textbf{81} (2010)
		127301, [arXiv:1005.3039].
		\bibitem {ref26}
		G. Kofinas and E. N. Saridakis, Teleparallel equivalent of Gauss-Bonnet gravity and its	modifications, \textit{Phys. Rev. D} \textbf{90}, 084044 (2014) [arXiv:1404.2249].
		\bibitem {ref27}
		C.-Q. Geng, C.-C. Lee, E. N. Saridakis and Y.-P. Wu, Teleparallel dark energy, \textit{Phys. Lett. B} \textbf{704}
		(2011) 384–387, [arXiv:1109.1092].
		\bibitem {ref28}
		M. Hohmann, L. J\"{a}rv and U. Ualikhanova, Covariant formulation of scalar-torsion gravity, \textit{Phys. Rev. D} \textbf{97}, 104011 (2018) [arXiv:1801.05786].
		\bibitem {ref29}
		R. Aldrovandi and J. G. Pereira, Teleparallel Gravity: An Introduction, Springer, Dordrecht (2013).
		\bibitem {ref30}
		J. W. Maluf, The teleparallel equivalent of general relativity, \textit{Annalen Phys.} \textbf{525}, (2013) 339,
		[arXiv:1303.3897].
		\bibitem {ref31}
		F. W. Hehl, J. D. McCrea, E. W. Mielke and Y. Ne'eman, Metric affine gauge theory of gravity: Field equations, Noether identities, world spinors, and breaking of dilation invariance, \textit{Phys. Rept.} \textbf{258}, 1 (1995) [arXiv:gr-qc/9402012].
		\bibitem {ref32}
		J. Beltran Jimenez, A. Golovnev, M. Karciauskas and T. S. Koivisto, The Bimetric variational principle for General Relativity, \textit{Phys. Rev. D} \textbf{86}, 084024 (2012) [arXiv:1201.4018].
		\bibitem {ref33}
		N. Tamanini, Variational approach to gravitational theories with two independent connections, \textit{Phys. Rev. D} \textbf{86}, 024004 (2012) [arXiv:1205.2511].
		\bibitem {ref34}
		G. Y. Bogoslovsky and H. F. Goenner, Finslerian spaces possessing local relativistic symmetry, \textit{Gen. Rel. Grav.} \textbf{31}, 1565 (1999) [arXiv:gr-qc/9904081].
		\bibitem {ref35}
		N. E. Mavromatos, S. Sarkar and A. Vergou, Stringy Space-Time Foam, Finsler-like Metrics and Dark Matter Relics, \textit{Phys. Lett. B} \textbf{696}, 300 (2011) [arXiv:1009.2880].
		\bibitem {ref36}
		S. Basilakos, A. P. Kouretsis, E. N. Saridakis and P. Stavrinos, Resembling dark energy and modified gravity with Finsler-Randers cosmology, \textit{Phys. Rev. D} \textbf{88}, 123510 (2013) [arXiv:1311.5915].
		\bibitem {ref37}
		A. P. Kouretsis, M. Stathakopoulos and P. C. Stavrinos, Covariant kinematics and gravitational bounce in Finsler space-times, \textit{Phys. Rev. D} \textbf{86}, 124025 (2012) [arXiv:1208.1673].
		\bibitem {ref38}
		A. Triantafyllopoulos and P. C. Stavrinos, Weak field equations and generalized FRW cosmology on the tangent Lorentz bundle, \textit{Class. Quant. Grav.} \textbf{35} 085011 (2018).
		\bibitem {ref39}
		S. Ikeda, E. N. Saridakis, P. C. Stavrinos and A. Triantafyllopoulos, Cosmology of Lorentz fiber-bundle induced scalar-tensor theories, \textit{Phys. Rev. D} \textbf{100} 124035 (2019) [arXiv:1907.10950].
		\bibitem {ref40}
		A. Conroy and T. Koivisto, The spectrum of symmetric teleparallel gravity, \textit{Eur. Phys. J. C} \textbf{78} \textbf{923} (2018) [arXiv:1710.05708].
		\bibitem {ref41}
		R. Myrzakulov, FRW Cosmology in $F(R,T)$ gravity, \textit{Eur. Phys. J. C} \textbf{72}, 2203 (2012)	[arXiv:1207.1039].
		\bibitem {ref42}
		E. N. Saridakis, S. Myrzakul, K. Myrzakulov and K. Yerzhanov, Cosmological applications of $F(R, T)$ gravity with dynamical curvature and torsion, \textit{Phys. Rev. D} \textbf{102} 023525 (2020) [arXiv:1912.03882].
		\bibitem {ref43}
		M. Jamil, D. Momeni, M. Raza and R. Myrzakulov, Reconstruction of some cosmological models in $f(R,T)$ gravity, \textit{Eur. Phys. J. C} \textbf{72}, 1999 (2012) [arXiv:1107.5807].
		\bibitem {ref44}
		M. Sharif, S. Rani and R. Myrzakulov, Analysis of $F(R, T)$ gravity models through energy conditions, \textit{Eur. Phys. J. Plus} \textbf{128}, 123 (2013) [arXiv:1210.2714].
		\bibitem {ref45}
		S. Capozziello, M. De Laurentis and R. Myrzakulov, Noether Symmetry Approach for teleparallel-curvature cosmology, \textit{Int. J. Geom. Meth. Mod. Phys.} \textbf{12} 1550095 (2015) [arXiv:1412.1471].
		\bibitem {ref46}
		P. Feola, X. J. Forteza, S. Capozziello, R. Cianci and S. Vignolo, The mass-radius relation for neutron stars in $f(R) = R + \alpha R^{2}$ gravity: a comparison between purely metric and torsion formulations, (2019) [arXiv:1909.08847].		
		\bibitem {ref47}
		F.K. Anagnostopoulos, S. Basilakos, E.N. Saridakis, Observational constraints on Myrzakulov gravity, (2020). [arXiv:2012.06524].
		\bibitem {ref48}
		N. Myrzakulov, R. Myrzakulov, L. Ravera, Metric-Affine Myrzakulov Gravity Theories, (2021). [arXiv:2108.00957].
		\bibitem {ref49}
		D. Iosifidis, N. Myrzakulov, R. Myrzakulov, Metric-Affine Version of Myrzakulov $F(R,T,Q,T)$ Gravity and Cosmological Applications, \textit{Universe} \textbf{7} 262 (2021). [ arXiv:2106.05083 ]
		\bibitem {ref50}
		T. Harko, N. Myrzakulov, R. Myrzakulov, S. Shahidi, Non-minimal geometry-matter couplings in Weyl-Cartan space-times: Myrzakulov $F(R,T,Q,T_{m})$ gravity (2022). [arxiv:2110.00358v1].
		\bibitem {ref51}
		R. Saleem, Aqsa Saleem, Variable constraints on some Myrzakulov models to study Baryon asymmetry, \textit{Chinese Journal of Physics} \textbf{84} 471-485 (2023).
		\bibitem {ref52}
		D. Iosifidis, R. Myrzakulov, L. Ravera, G. Yergaliyeva, K. Yerzhanov, Metric-Affine Vector-Tensor Correspondence and Implications in $F(R,T,Q,T, D)$ Gravity (2021). [ arXiv:2111.14214].
		\bibitem {ref53}
		G. Papagiannopoulos, S. Basilakos, E.N. Saridakis, Dynamical system analysis of Myrzakulov gravity, (2022). [arXiv:2202.10871]
		\bibitem {ref54}
		S. Kazempour, A. R. Akbarieh, Cosmological Study in $F(R, T)$ Quasi-dilaton Massive Gravity, (2023). [arXiv:2309.09230].
		\bibitem {ref55}
		F. K. Anagnostopoulos, S. Basilakos, E. N. Saridakis, Observational constraints on Myrzakulov gravity, \textit{Phys. Rev. D} \textbf{103}, 104013 (2021). [arXiv:2012.06524 [gr-qc]].
		\bibitem {ref56}
		D. C. Maurya, R. Myrzakulov, Transit cosmological models in Myrzakulov $F(R,T)$ gravity theory, (2024) [arXiv:2401.00686 [gr-qc]].
		\bibitem {ref57}
		D.C. Maurya, A. Dixit, and A Pradhan, Transit string dark energy models in $f(Q)$ gravity, \textit{Inter. J. Geom. Meth. Mod. Phys.} \textbf{20} 2350134 (2023).
		\bibitem {ref58}
		D.C. Maurya, Phantom Dark Energy Nature of String-Fluid Cosmological Models in $f(Q)$-Gravity, \textit{Gravitation and Cosmology} \textbf{29} (4), 345-361 (2023).
		\bibitem {ref59}
		D.C. Maurya and J. Singh, Modified $f(Q)$-Gravity String Cosmological Models With Observational Constraints, \textit{Astronomy and Computing} \textbf{46} 100789 (2024). https://doi.org/10.1016/j.ascom.2024.100789.
		\bibitem {ref60}
		D.C. Maurya, Reconstructing $\Lambda$CDM $f(T)$ gravity model with observational constraints, \textit{Inter. J. Geom. Meth. Mod. Phys.}, (2024) 2450039, https://doi.org/10.1142/S0219887824500397.
		\bibitem {ref61}
		A. Dixit, A. Pradhan, and D.C. Maurya, A probe of cosmological models in modified teleparallel gravity, \textit{Inter. J. Geom. Meth. Mod. Phys.} \textbf{18} 2150208 (2023).
		\bibitem {ref62}
		D.C. Maurya, Accelerating scenarios of viscous fluid universe in modified $f(T)$ gravity, \textit{Inter. J. Geom. Meth. Mod. Phys.} \textbf{19} 2250144 (2022).
		\bibitem {ref63}
		R Zia, DC Maurya, and AK Shukla, Transit cosmological models in modified $f(Q,T)$ gravity, \textit{Inter. J. Geom. Meth. Mod. Phys.} \textbf{18}, 2150051 (2021).
		\bibitem {ref64}
		A. Paliathanasis, S. Basilakos, E. N. Saridakis, S. Capozziello, K. Atazadeh, F. Darabi and	M. Tsamparlis, New Schwarzschild-like solutions in $f(T)$ gravity through Noether symmetries, \textit{Phys. Rev. D} \textbf{89}, 104042 (2014) [arXiv:1402.5935].
		\bibitem {ref65}
		A. Paliathanasis, $f(R)$-gravity from Killing Tensors, \textit{Class. Quant. Grav.} \textbf{33} 075012 (2016)			[arXiv:1512.03239].
		\bibitem {ref66}
		N. Dimakis, A. Karagiorgos, A. Zampeli, A. Paliathanasis, T. Christodoulakis and P. A. Terzis, General Analytic Solutions of Scalar Field Cosmology with Arbitrary Potential, \textit{Phys. Rev. D} \textbf{93} 123518 (2016) [arXiv:1604.05168].
		\bibitem {ref67}
		V. Sahni, \textit{et al.}, Statefinder-a new geometrical diagnostic of dark energy, \textit{JETP Lett.} \textbf{77}, 201 (2003).
		\bibitem {ref68}
		U. Alam, \textit{et al.}, Exploring the expanding universe and dark energy using the Statefinder diagnostic, \textit{Mon. Not. R. Astron. Soc.} \textbf{344}, 1057 (2003).
		\bibitem {ref69}
		M. Sami,\textit{ et al.}, Cosmological dynamics of a nonminimally coupled scalar field system and its late time cosmic relevance, \textit{Phys. Rev. D} \textbf{86}, 103532 (2012).
		\bibitem {ref70}
		D.W. Hogg and D.F. Mackey, Data analysis recipes: Using Markov Chain Monte Carlo, \textit{The Astrophysical Journal Supplement Series} \textbf{236} (2018) 18. arXiv:1710.06068 [astro-ph.IM].
		\bibitem {ref71}
		C. Zhang, \textit{et al.}, Four new observational $H(z)$ data from luminous red galaxies in the Sloan Digital Sky Survey data release seven, \textit{Research in Astronomy and Astrophysics}, \textbf{14} 1221 (2014).
		\bibitem {ref72}
		J. Simon \textit{et al.}, Constraints on the redshift dependence of the dark energy potential, \textit{Phys. Rev. D} \textbf{71} 123001 (2005).
		\bibitem {ref73}
		M. Moresco, \textit{et al.}, Improved constraints on the expansion rate of the Universe up to $z\sim1.1$ from the spectroscopic evolution of cosmic chronometers, \textit{J. Cosmology Astropart. Phys.}, \textbf{8}, 006 (2012).
		\bibitem {ref74}
		M. Moresco, \textit{et al.}, A $6\%$ measurement of the Hubble parameter at $z\sim0.45$: direct evidence of the epoch of cosmic re-acceleration, \textit{J. Cosmology Astropart. Phys.}, \textbf{5}, 014 (2016).
		\bibitem {ref75}
		A. L. Ratsimbazafy, \textit{et al.}, Age-dating luminous red galaxies observed with the Southern African Large Telescope, \textit{MNRAS}, \textbf{467}, 3239 (2017).
		\bibitem {ref76}
		D. Stern, \textit{et al.}, Cosmic chronometers: constraining the equation of state of dark energy. I: $H(z)$ measurements, \textit{J. Cosmology Astropart. Phys.}, \textbf{2}, 008 (2010).
		\bibitem {ref77}
		N. Borghi, \textit{et al.}, Toward a Better Understanding of Cosmic Chronometers: A New Measurement of $H(z)$ at $z\sim0.7$, \textit{Astrophys. J. Lett.} \textbf{928}, L4 (2022).
		\bibitem {ref78}
		M. Moresco, Raising the bar: new constraints on the Hubble parameter with cosmic chronometers at $z\sim2$, \textit{MNRAS}, \textbf{450}, L16 (2015).
		\bibitem {ref79}
		D. M. Scolnic \textit{et al.}, The complete light-curve sample of spectroscopically confirmed SNe Ia from Pan$-$STARRS1 and cosmological constraints from the combined pantheon sample, \textit{Astrophys. J.} \textbf{859} (2018) 101.
		\bibitem {ref80}
		S. Cao and B. Ratra, $H_{0}=69.8\pm1.3~km~s^{-1}~Mpc^{-1}$, $\Omega_{m0}=0.288\pm0.017$, and other constraints from lower-redshift, non-CMB, expansion-rate data, \textit{Phys. Rev. D} \textbf{107}, 103521 (2023). [arXiv:2302.14203 [astro-ph.CO]].
		\bibitem {ref81}
		S. Cao and B. Ratra, Using lower-redshift, non-CMB, data to constrain the Hubble constant and other cosmological parameters, \textit{MNRAS} \textbf{513}, 5686-5700 (2022). [arXiv:2203.10825 [astro-ph.CO]].
		\bibitem {ref82}
		Omer Farooq, Bharat Ratra, Hubble parameter measurement constraints on the cosmological deceleration-acceleration transition redshift, (2013) [ arXiv:1301.5243v1 [astro-ph.CO] ].
		\bibitem {ref83}
		Omer Farooq, Sara Crandall, Bharat Ratra, Binned Hubble parameter measurements and the cosmological deceleration-acceleration transition, (2013) [ arXiv:1305.1957v1 [astro-ph.CO]].
		\bibitem {ref84}
		Omer Farooq, Foram Madiyar, Sara Crandall, Bharat Ratra, Hubble parameter measurement constraints on the redshift of the deceleration-acceleration transition, dynamical dark energy, and space curvature, (2016) [arXiv:1607.03537v2 [astro-ph.CO] ].
		\bibitem {ref85}
		Hai Yu, Bharat Ratra, Fa-Yin Wang, Hubble Parameter and Baryon Acoustic Oscillation Measurement Constraints on the Hubble Constant, the Deviation from the Spatially-Flat $\Lambda$cdm Model, The Deceleration-Acceleration Transition Redshift, and Spatial Curvature, (2018) [arXiv:1711.03437v2 [astro-ph.CO] ].
		\bibitem {ref86}	
		D.C. Maurya, R. Zia, Brans-Dicke scalar field cosmological model in Lyra’s geometry, \textit{Physical Review D} \textbf{100}(2) 023503 (2019). https://doi.org/10.1103/PhysRevD.100.023503.
		\bibitem {ref87}
		D.C. Maurya, J. singh, L.K. Gaur, Dark Energy Nature in Logarithmic $f(R,T)$ Cosmology, \textit{Inter. J. Geom. Meth. Mod. Phys.} \textbf{20}(11) 2350192 (2023). https://doi.org/10.1142/S021988782350192X.
	\end{thebibliography}
\end{document}